\documentclass[usenatbib,usedcolumn,onecolumn]{mnras}
\pdfoutput=1

\usepackage[T1]{fontenc}
\usepackage{ae,aecompl}
\usepackage[british]{babel}             
\let\la=\lesssim 

\usepackage{graphicx,url,booktabs,xspace,lastpage}
\usepackage{amsmath,amssymb}
\usepackage{mathtools, cuted}
\usepackage{verbatim}
\usepackage{color, colortbl}
\definecolor{Gray}{gray}{0.9}

\usepackage[normalem]{ulem} 
\newcommand{\strike}{\bgroup\markoverwith{\textcolor{red}{\rule[0.5ex]{2pt}{0.4pt}}}\ULon}
\newcommand{\cfsout}{\bgroup\markoverwith{\textcolor{red}{\rule[0.5ex]{2pt}{0.4pt}}}\ULon}

\AtBeginDocument{\colorlet{defaultcolor}{.}}
\usepackage{color,xcolor}
\definecolor{amethyst}{rgb}{0.6, 0.4, 0.8}
\definecolor{green}{rgb}{0.55, 0.71, 0.0}
\newcommand{\re}[1]{\textcolor{defaultcolor}{#1}} 
\newcommand{\ree}[1]{\textcolor{defaultcolor}{#1}} 

\hypersetup{pdfauthor={},
               pdftitle={},
               pdfkeywords={instrumentation: adaptive optics,site testing,telescopes,atmospheric effects},
               bookmarksnumbered=true}

\newcommand{\avg}[1]{\left< #1 \right>} 
\newcommand{\ud}{\mathrm{d}}
\newcommand{\thetam}{\theta^\mathrm{max}_\mathrm{tel}}

\newcommand{\thetal}{\theta_\mathrm{lgs}}
\newcommand{\phil}{\phi_\mathrm{lgs}}
\newcommand{\Hm}{H_\mathrm{max}}
\newcommand{\Hmol}{H_\mathrm{mol}}
\newcommand{\Haer}{H_\mathrm{aer}}
\newcommand{\htr}{h_\mathrm{track}}
\newcommand{\hcta}{h_\mathrm{CTA}}
\newcommand\ddfrac[2]{\frac{\displaystyle #1}{\displaystyle #2}}
\newcommand{\ns}{N_\mathrm{s}}
\newcommand{\betam}{\beta_\mathrm{mol}}
\newcommand{\betaa}{\beta_\mathrm{aer}}
\newcommand{\fovp}{\textit{FOV}_\mathrm{pix}}
\newcommand{\fovc}{\textit{FOV}_\mathrm{camera}}
\newcommand{\pde}{\textit{PDE}_{589\,\mathrm{nm}}}
\newcommand{\atel}{A_\mathrm{tel}}
\newcommand{\pveto}{P_\mathrm{veto}}
\newcommand{\pobs}{P_\mathrm{obs}}

\newcommand{\rpix}{R_\mathrm{pix}}
\newcommand{\rcrit}{R_\mathrm{crit}}
\newcommand{\nlas}{N_\mathrm{lasers}}
\newcommand{\dtrack}{d_\mathrm{track}}
\newcommand{\Dmin}{D_\mathrm{min}}
\newcommand{\Dmax}{D_\mathrm{max}}
\newcommand{\tair}{T_\mathrm{air}}

\DeclareMathOperator\atan{atan}

\title[Impact of LGS system on CTA]{Impact of Laser Guide Star
facilities on neighbouring telescopes:  The case of GTC, TMT, \re{VLT} and \re{ELT} lasers and the Cherenkov Telescope Array}

\author[M. Gaug et al.]{
M. Gaug,$^{1}$\thanks{E-mail: markus.gaug@uab.cat}
M. Doro$^{2}$
\\
$^{1}$Unitat de F\'isica de les Radiacions, Departament de F\'isica, and CERES-IEEC, \\Universitat Aut\`onoma de Barcelona, E-08193 Bellaterra, Spain\\
$^{2}$Department of Physics and Astronomy (DFA) of the University of Padova \& INFN Padova, I-35131 Padova, Italy\\
}

\date{Accepted XXX. Received YYY; in original form ZZZ}

\pubyear{2018}

\volume{{\rm in press}}

\begin{document}
\label{firstpage}
\pagerange{\pageref{firstpage}--\pageref{LastPage}}
\maketitle

\begin{abstract}
  Powerful Laser Guide Star (LGS) systems are standard for the next-generation of extremely large telescopes.
  However, modern earth-based astronomy has gone through a process of concentration on few sites with exceptional sky quality,
  resulting in those becoming more and more crowded. The future LGS systems encounter hence an environment of surrounding
  astronomical installations, some of which observing with large fields-of-view. We derive formulae to calculate the impact of
  LGS light on the camera of a neighbouring telescope and the probabilities for a laser crossing the camera field-of-view to occur,  
  and apply these to the specific case of the next very-high-energy gamma-ray observatory ``Cherenkov Telescope Array'' (CTA).
  Its southern part shall be constructed in a valley of the Cerro Armazones, Chile,
  close to \re{the ``Very Large Telescope'' (VLT) and} the ``European Extremely Large Telescope'' \re{(ELT)}, 
  while its northern part will be located at the ``Observatorio del Roque de los Muchachos'', on the Canary Island 
  of La~Palma, which also hosts the ``Gran Telescopio de Canarias'' (GTC)
  and serves as an optional site for the ``Thirty Meter Telescope'' (TMT), both employing LGS systems. 
  Although finding the artificial star in the field-of-view of
  a CTA telescope will not disturb observations considerably,
  the laser beam crossing the field-of-view of a CTA telescope may be critical. 
  We find no conflict expected for the \re{ELT} lasers, 
  however, 1\% (3\%) of extra-galactic and  1\% (5\%) of galactic observations with the CTA
    may be affected by the GTC (TMT) LGS lasers,
  unless an enhanced version of a laser tracking control system gets implemented.
\end{abstract}


\section{Introduction}

Laser Guide Stars (LGS) systems~\citep{Bonaccini:2010,2014AdOT....3..345C,orgeville:2016} provide artificial 
  reference sources to partially correct the impact of atmospheric
  turbulence on astronomical observations. They are used in
  coincidence with Adaptive Optics (AO) systems. LGS are used to
  provide increased sky coverage and availability compared to natural
  guide stars~\citep{Foy:1985}. 
%
  \re{Mostly}, high power lasers tuned to the \re{D$_{2a}$} 
  resonance of sodium atoms (at \re{589.159}~nm \ree{in vacuum}) are \re{propagated} at a sky location
  within the field of view of the optical telescope for which the
  wavefront needs correction. 
  High power sodium lasers produce artificial stars 
  by exciting a layer of sodium atoms \re{from their 3S$_\mathrm{1/2}$ to the 3P$_\mathrm{3/2}$ level} in the mesosphere \re{which produce fluorescence emission while de-exciting.}
  \re{The emission is centered at an altitude of $(91.9 \pm 0.8)$~km a.s.l. and has an equivalent full width at half maximum of $\sim (11.3\pm 1.2)$~km~\citep{Moussaoui:2010}.}
The \re{used laser light is often} circularly polarized to achieve maximum impact~\citep{Holzloehner:2010,Boyer:2010}.  
The creation of several guide stars is also possible, to achieve asterism with a \re{radial distance from science target} ranging from  
\re{0.5$^{\prime}$ to 6$^{\prime}$} on the sky. 
The LGS will likely be operated regularly during observations, and their
 scattered light (Rayleigh and Mie) will then be seen by other telescopes until distances of several kilometers from the location of their host observatory.
 Assuming the close-by installation 
 observes in a wavelength range enclosing that of the LGS lasers,
 the scattered laser light may then leave spurious light tracks on 
 the cameras and affect operation in several ways: $a)$ by generating false
 triggers \re{(for installations which trigger image readout e.g. from Cherenkov light pulses)}
 the star guider camera and the precision pointing of the telescopes,  
 and ultimately, $d)$ by affecting the duty cycle, if active laser avoidance is chosen.
 Several of the enumerated problems can be often overcome with the use of Notch-filters~\citep{Schallenberg:2010} or band-pass filters~\citep{magicmoon,veritasmoon}, however this is not always possible at a reasonable cost,
 particularly not in the case of the CTA, where every camera pixel would need to be covered by such a filter. Light losses at smaller wavelengths, particularly in the sensitive region from 300~nm to 500~nm 
 need to be strictly controlled in order to ensure that sensitivity losses remain acceptable, particularly around the energy threshold of the CTA.
 It is therefore important to compute the amount of light that can reach a neighbouring installation, as well as to discuss the  probability of interferences.
 The latter depends on the angular separation between
 the direction of the lasers and the telescopes' optical axis, the
 distance to the crossing point and its altitude, the size of the
 collecting surface of the telescopes, and the photon detection
 efficiency (PDE) of the photo-sensors. 
The goal of this study is that to provide a reference formalism to
address this twofold interference (spurious light yield and
probability of crossings). This is done through the paper with a
general approach, but is quantified for the particular case of the
oncoming Cherenkov Telescope Array (CTA).

CTA~\citep{Actis:2011} will be an
observatory for   gamma-ray astronomy in the GeV-TeV energy range. It
is based on the so-called Imaging Atmospheric Cherenkov Technique
(IACT) that captures the Cherenkov light emitted by extensive air
showers (EAS), produced when very high energy gamma-rays hit
the Earth's atmosphere. The EAS is a cascade of a large number of
sub-atomic particles (mainly electrons and positrons)  which reaches a maximum at
altitudes between 8--12~km a.s.l. for a 1~TeV shower, on average, however, moving to considerably higher altitudes at lower energies.
The  relativistic particles forming the cascade emit Cherenkov light which propagates towards the ground. The Cherenkov light emission is
strongest in the ultraviolet and blue, hence the CTA telescopes and cameras are
optimized to a wavelength range \re{from 300~nm to 500~nm}, but are  
also sensitive in the green, and even yellow, part of the optical
spectrum. Silicon-photomultiplier-based cameras may even extend sensitivity \re{beyond 900~nm~\citep{Otte:2017}}.

\begin{figure}
  \centering
  \includegraphics[width=0.45\linewidth]{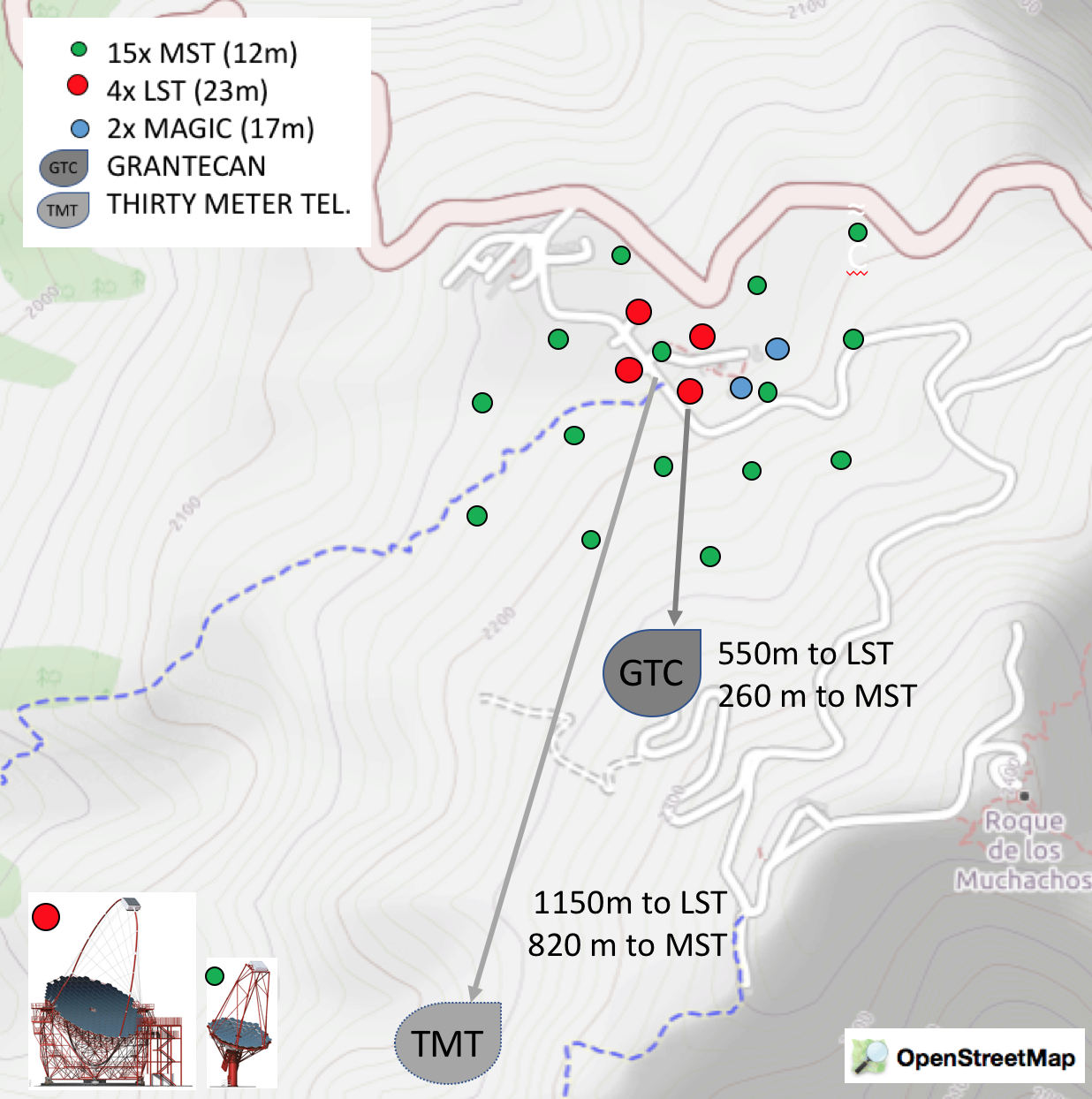}
  \includegraphics[width=0.45\linewidth]{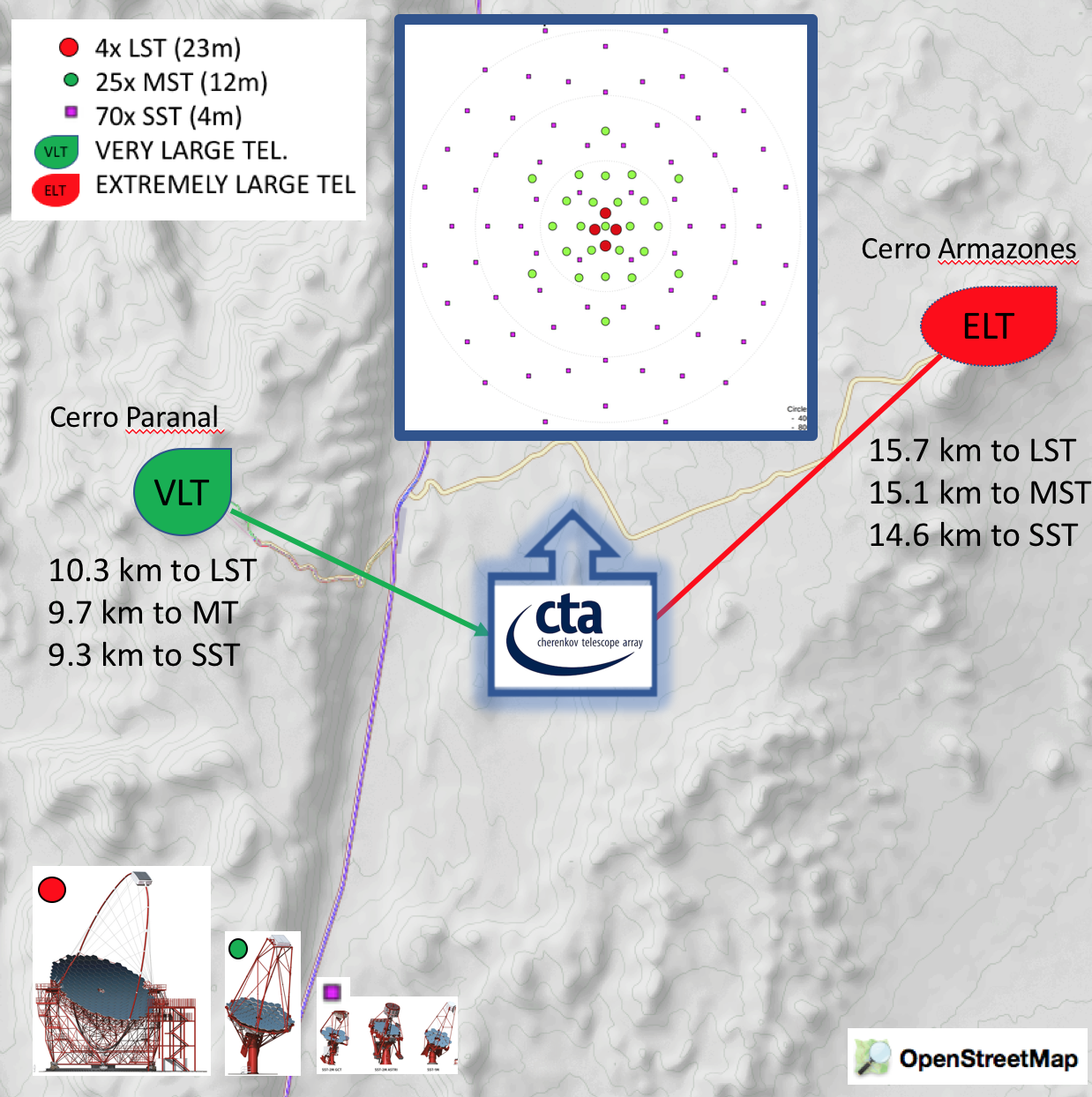}
  \caption{\label{fig:cta_aray}\re{Schematic view of the Northern
    Hemisphere Site of CTA (CTA-N, left) at the \textit{Observatorio del Roque de
    Los Muchachos} on the Canary Island of La Palma, Spain, and of the Southern CTA Site
    (CTA-S, right) at the ESO site of \textit{Cerro Armazones}, Chile. 
    The planned disposition of CTA telescopes is shown, together with the
    locations of the existing VLT and GTC telescopes, the ELT under
    construction as well as the
    possible location of the TMT. Distances from the optical
    telescopes to the closest CTA telescopes are marked. The underground map
    has been obtained with \texttt{openstreetmap.org} and the CTA
    layout from \texttt{www.cta-observatory.org}}.}
\end{figure}

CTA will operate at two sites: one in the Northern Hemisphere, at the
Observatorio del Roque de los Muchachos (ORM), La Palma, Spain, and
one in the Southern Hemisphere, at a Chilean site 
of the European Southern Observatory (ESO), close to Paranal. The northern array
(hereafter CTA-N) has been formally accepted, and construction has already started. Negotiations about the
southern array (hereafter CTA-S) are close to being concluded with the
Chilean authorities and the ESO.
Both arrays  will consist of several telescopes of different
  sizes: four Large-Sized-Telescopes (LST) of 23~m diameter mirror
dish each, in the central core of the array, 15 Medium-Sized-Telescopes (MST) of a 12~m
diameter mirror in CTA-N and 25 MSTs in CTA-S surrounding the LSTs, while CTA-S will be additionally equipped
with 70 Small-Sized-Telescopes (SST) of an equivalent mirror diameter of about 4~m\footnote{%
see also \url{https://www.cta-observatory.org} }.
In the Southern array, telescopes will be installed across a large area
of roughly $2.2 \times 2.4$~km, centered at 24.674$^\circ$\,S,
70.316$^\circ$\,W, at 2150~m a.s.l. At a distance of about \re{10~km
in the NW direction} from the
center of CTA-S, \re{the Very Large Telescope
(VLT)}\footnote{\url{www.eso.org/public/teles-instr/paranal-observatory/vlt/}}
\re{is taking data since 1998. Slightly further away, in the NE direction,} at 15.8~km, the
European Extremely Large Telescope
\re{(ELT)}\footnote{\url{https://www.eso.org/sci/facilities/eelt/}} is 
under construction. The Northern array is somewhat smaller, covering an area of about $500
\times 600$~m,   
between the current ORM residence and the higher altitude rim of the Caldera de
Taburiente Mountain, which hosts several optical telescopes. In between, the Gran Telescopio de Canarias
(GTC)\footnote{\url{http://www.gtc.iac.es/}}, located at 28.762$^\circ$\,N, 
17.892$^\circ$\,W, is found, at about 550~m from the center of the
CTA-N. Further in that direction, at a distance of 1150~m from the
center of the CTA-N {and located at 28.753$^\circ$\,N,
  17.897$^\circ$\,W, the Thirty-Meter Telescope (TMT)\footnote{\url{https://www.tmt.org/}} may be installed, if finally the ORM is chosen to host it. 
  Because the MST and SST telescopes are wider distributed around the central LSTs, several of them will  in some
  cases further approach the \re{VLT or} \re{ELT} (at the CTA-S), or the GTC or TMT (at the 
  CTA-N).  \re{See Figure~\ref{fig:cta_aray} for a schematic view of both sites and their neighbouring installations.}
The \re{four} above-mentioned telescopes \re{VLT}, \re{ELT}, GTC, and TMT, 
\re{incorporate or} will incorporate Laser Guide Star Facilities (LGSF) which contain powerful
continuous wave~lasers~\citep{Ageorges:2000,wei:2012,Herriot:2014SPIE} to create artificial guide-stars for the adaptive optics
(AO) system of their primary mirrors: one in the case of
the GTC, \re{four in the case of VLT} and  up to six in the case of the extremely large
telescopes \re{ELT} and TMT.
The TMT plans to create some asterism such
that the outermost laser-guide-star  will be distributed along circles
with a perimeter of typically between 35$^{\prime\prime}$ and 70$^{\prime\prime}$~\citep{Boyer:2010}\footnote{%
Larger asterisms reaching up to 510$^{\prime\prime}$ in perimeter have been presented in \citet{Boyer:2010}, but require more than 6 LGS, which are currently not foreseen.}.
Each AO-laser itself is extremely well
collimated ($\simeq O(\mathrm{arcsec})$) and operates at \ree{the vacuum} wavelength
of \re{$\lambda_\mathrm{lgs}$ = 589.159}~nm,  with a typical exit power of order $\sim$20~W, after exiting the
beam transfer optics. \re{Systems operating pulsed lasers in the UV (355~nm)~\citep{Tokovinin:2016} and in green (515 or 532~nm)~\citep{Rutten:2006,Rabien:2011} have been built as well, but are not a primary option for the extremely large telescopes. }
\re{The LGS} lasers will operate
till elevations of as low as 20$^\circ$ and could therefore cross the
view-cone of some of the telescopes of the CTA.

  
\bigskip
The paper is structured as follows: in \autoref{sec:computation},
  we compute the amount of LGS induced light on a generic camera receiver unit
  pixel at the position of a putative neighbouring instrument. In
  \autoref{sec.prob}, we estimate the probabilities that the LGS laser
  light beams cross the field-of-view of a neighbouring instrument and impede re-positioning. In
  \autoref{sec:results} we quantify the two effects for the realistic
  case of CTA telescopes in each hemisphere and the planned close-by LGS
  facilities. In \autoref{sec:discussion} we discuss the
  results and conclude. There follow some appendices for the derivation of several larger formulae.


\section{Computation of AO-laser induced light on neighbouring installations}
\label{sec:computation}

\begin{table*}
\centering
\begin{tabular}{lllllp{4.5cm}}
  \toprule
    \rowcolor{Gray}
Parameter  &  Value & Value  & Value   &  Value & Comments \\
  \rowcolor{Gray}
&  (GTC) &  (TMT) & \re{(ELT)} & \re{(VLT)} &  \\
\midrule
Number of lasers & 1    & 6  & 6   &  \re{4} &  \\ 
DC power         & 16.5~W & $6 \times 16.5$~W & $6\times 16.5$~W & \re{$4\times 17$~W} & 22~W  laser, assuming 75\% Beam Transfer Optics and Laser Launch Telescopes throughput~\citep{Bonaccini:2010}, 88\% for VLT~\citep{2014AdOT....3..345C} \\
\ree{Vacuum} wavelength       &   \re{589.159~nm} & \re{589.159~nm}  &  \re{589.159~nm}  & \re{589.159~nm}  & \re{minor admixtures of 589.157~nm and 589.611~nm~\citep{Vogt:2017}}  \\
Operation elevation  & 30$^\circ$--90$^\circ$ & 25$^\circ$--90$^\circ$  & \re{20$^\circ$--90$^\circ$} &  \\
Duty Cycle AO-laser  & 15\% & $\gtrsim$75\% &  \re{$\sim 50$\% } & \\
Latitude             & $28^\circ$45$^\prime$23.8$^{\prime\prime}$~N & $28^\circ$45$^\prime$09$^{\prime\prime}$~N & $24^\circ$35$^\prime$21$^{\prime\prime}$~S  & \re{$24^\circ$37$^\prime$38$^{\prime\prime}$~S}  &  \\
Longitude            & $17^\circ$53$^\prime$30.8$^{\prime\prime}$~W & $17^\circ$53$^\prime$45$^{\prime\prime}$~W & $70^\circ$11$^\prime$39$^{\prime\prime}$~W  & \re{$70^\circ$24$^\prime$15$^{\prime\prime}$~W}  & \\
Altitude             & 2280~m a.s.l. & 2300~m a.s.l. &  3050~m a.s.l. & \re{2640~m a.s.l.} &  \\
\midrule
Closest distance to LST & 0.55~km & 1.15~km & 15.7~km & \re{10.3~km} & \\
Closest distance to MST & 0.26~km & 0.82~km & 15.1~km & \re{9.7~km} & \\
Closest distance to SST & --      & --      & 14.6~km & \re{9.3~km} &  \\
\midrule
Altitude difference to LST & 55~m  & 75~m & 910~m  & \re{500~m} & \\
Altitude difference to MST & 50~m  & 40~m & 900~m  & \re{480~m} & only for closest MST \\
Altitude difference to SST & -- & -- & 870~m       & \re{450~m} & only for closest SST \\
\bottomrule
\end{tabular}
\caption{\label{tab:laser_params} Characteristics of the AO~laser
  systems of the GTC, TMT, VLT and \re{ELT} telescopes and distances compared to CTA telescopes. }
\end{table*}

Throughout this section, the number of photons produced in photon
  detection systems of another instrument far from the LGS system is
  computed. Quantitative numbers are computed for the CTA,
  however, the formulae are kept as general as possible, and can be applied to any
  similar case.

The main parameters of the LGS systems of VLT, \re{ELT}, GTC and TMT used for
this study are reported in \autoref{tab:laser_params} together with 
their closest distance to CTA telescopes and altitude differences.
Based on the experience with the 4 Laser Guide Star Facility (4LGSF) on Unit Telescope 4 (UT4) of \re{the VLT}~\citep{Vogt:2017}, 
an LGS system will be implemented on the \re{ELT}~\citep{Fusco:2010}, which will rely on continuous wave lasers similar to those of the 4LGSF (according to the 
current design)~\citep{Bonaccini:2010}.
The TMT will operate six
lasers, each with a power of 22~W (16.5~W after
exiting the beam transfer optics)~\citep{Herriot:2014SPIE}.
%
Finally, the upgrade of the GTC AO system with an LGS facility has been recenty approved and the system is now entering its conceptual design phase~\citep{Garcia-Talavera:2016a,GTCAO:website}.

\bigskip
To compute the effect of the AO-laser light on the CTA camera pixels, we consider first
typical scattering scenarios in the lower atmosphere, and their formulation in the framework of Rayleigh and Mie scattering
(\autoref{sec:scattering})
and secondly, derive an equation for the amount of light imaged into one camera pixel, considering the geometry of the problem (\autoref{sec:geometry}). We
discuss two case scenarios in a later section (\autoref{sec:scenarios}).

\subsection{Scattering of the laser light in the lower atmosphere}
\label{sec:scattering}

Light of wavelength $\lambda$ and polarization angle $\phi$ is scattered in the atmosphere by air molecules (through Rayleigh scattering) 
and aerosols (through Mie scattering, or even more complicated ways if the shape of the scattering aerosols is not radially symmetric~\citep{dubovik}). 
In dry air, light is \re{elastically}\footnote{%
\re{Additionally, Raman scattering on nitrogen and oxygen molecules has been observed~\citep{Vogt:2017}, albeit with intensities more than three orders of magnitude lower than the pure elastically scattered return.}}
  } scattered by air molecules at a scattering angle $\theta$ with respect
to the impinging photon direction into a solid angle with a differential
  cross-section $\ud\sigma/\ud\Omega$~\citep{Penndorf:1957,bucholtz}: 
\begin{equation}\label{eq.rayleigh}
\frac{\ud\sigma(\phi,\theta,\lambda)}{\ud\Omega} = \frac{9\pi^2 \cdot (n^2(\lambda)-1)^2 }{\lambda^4 \cdot \ns^2 \cdot (n^2(\lambda)+2)^2} \cdot \left(\frac{6+3\rho}{6-7\rho}\right) 
\cdot \left( \frac{2+2\rho}{2+\rho} \right) \left( \sin^2(\phi) +  \big( \frac{1-\rho}{1+\rho} \big) \cdot \cos^2(\phi)\cos^2(\theta)  \right) \;.   
\end{equation}
%
%
Here, $\ns$ is the molecular concentration, $n(\lambda)$ the
refractive index and $\rho$ the de-polarization ratio of air. 
Because $(n^2(\lambda)-1)/(n^2(\lambda)+2)$ is proportional to $\ns$, \autoref{eq.rayleigh} is independent of 
 density (as well as temperature and pressure)~\citep{Bodhaine:1999}, and depends only on the components' mixture of air (which can be assumed constant throughout the troposphere and in time, except for the 
negligible contribution of CO$_2$). 
One can hence pick a reference condition for temperature and pressure ($T,P$), typically chosen as 
the US~standard atmosphere~\citep[$T_s=288.15$~K and
  $P_s=1013.25$~mbar]{USstandard76} which yields $\ns = 2.547\cdot 10^{25}$~m$^{-3}$. 
At \re{$\lambda = \lambda_\mathrm{lgs}/n \approx 589.0$~nm}, the AO-laser wavelength in air, the combination $(n^2-1)/(n+2)$ yields then $1.84\times 10^{-4}$~\citep{Peck:1972}.  
Finally, the so-called \textit{King factor} $(6+3\rho)/(6-7\rho)$  describes the effect of the
molecular anisotropy of air and amounts to about 1.048 at 589~nm~\citep{tomasi}.
%
%
%
The factors $(2+2\rho)/(2+\rho)\approx 1.01$ and $(1-\rho)/(1+\rho) \approx 0.95$ describe the \textit{Chandrasekhar correction}~\citep{bucholtz,Chandrasekhar}.
After multiplying with the number density of molecules at a given
altitude~$h$, we obtain the \emph{volume scattering coefficient}
$\betam(\lambda,\theta,\phi,h)$~\citep[see also][]{gaug2014}: 
\begin{align}
  \betam(589.2~\mathrm{nm},\theta,\phi,h) 
  & \approx 1.0 \times 10^{-6} \cdot \left( 0.95 \cdot \cos^2(\phi)\cos^2(\theta) + \sin^2(\phi) \right) 
  \cdot \frac{N(h)}{\ns}~\mathrm{m^{-1}\,sr^{-1}}  \nonumber\\
  & \approx 1.0 \times 10^{-6} \cdot \frac{ 0.95\cdot \cos^2(\theta) + 1 }{2}
  \cdot \frac{N(h)}{\ns}~\mathrm{m^{-1}\,sr^{-1}} \quad,  \label{eq.betam}
\end{align}
where  \textit{un-polarized light} or a \textit{circularly polarized light} beam has been assumed in the second line. 
The scattering probability becomes radially symmetric in such a case.
%
%
To estimate the dependency of $N(h)/N_s$ on altitude, we use a typical atmospheric winter condition at the ORM\footnote{The main results of this study are however unaffected by
    this assumption.}~\citep{Gaug:2017site} with:
\begin{equation}
\frac{N(h)}{\ns} = f(h) \cdot \exp\left(-\frac{h}{\Hmol}\right) \quad,
\end{equation}
where $h$ is the altitude a.s.l. of the scattering point, 
and $\Hmol \approx 9.5~\mathrm{km}$ the average density scale height of the local troposphere \re{(9.8~km for the central Summer months)}. The function $f(h)$ reproduces the slight modulation of density in the tropopause
and the stratosphere and can be modelled with the following average correction function\footnote{obtained from fits to NASA's NRLMSISE-00 density profiles (\url{https://ccmc.gsfc.nasa.gov/modelweb/models/nrlmsise00.php}).}:
\begin{equation}
f(h) \approx \left\{ \begin{array}{ll}
   0.8845+0.0426\cdot h - 0.004 \cdot h^2 + 6.1 \times 10^{-5} \cdot h^3 & \textrm{for~$h<18.4$~km} \\
   1.5917-0.061\cdot h  + 0.667 \cdot h^2 & \textrm{for~$h>18.4$~km} 
\end{array} \right. \label{eq.fh}
\end{equation}
Whereas \autoref{eq.betam} is precise to a few percent, the correction function \autoref{eq.fh} can show variations of more than 10\%, particularly in the tropopause. 

\bigskip
Aerosols scatter light more efficiently than molecules and usually
  less isotropically, due to their
larger sizes, although they are much less in number density. 
World-class astronomical observatories are however characterized by extremely low aerosol contamination on average. 
For instance, typical winter nights on La Palma show {\it aerosol optical thicknesses} (AOTs) of the ground layer of the order 
of only 0.02 at $\lambda=532$~nm, with extinction coefficients distributed exponentially with a scale height of around $\Haer \approx 500$~m~\citep{Gaug:2017site}, hence: 
\begin{equation}
\alpha(\htr) = \alpha_{0,532\,\mathrm{nm}} \cdot \exp(-\htr/\Haer) \quad,
\end{equation}
where $\alpha_{0,532\,\mathrm{nm}} \approx 4.5 \times 10^{-5}$~m$^{-1}$ \re{and $\htr$ is the altitude of the observed part of the laser track above the neighboring telescope}.
At Paranal, only the AOT has been studied so far~\citep{patat2011}, yielding similar results.

We further assume a typical
\AA ngstr\"om index in the range from 0.5--1.5 for clear nights~\citep[see e.g. entries ``IZA'' or ``MLO'' in Fig.~3 of][]{Andrews2011},
and derive $\alpha_{0,589\,\mathrm{nm}} \approx 4 \times 10^{-5}$~m$^{-1}$ for $\lambda = 589$~nm. 
As we will later see, this number becomes important only on rare occasions. 
We can use the \textit{Henyey-Greenstein} formula~\citep{Henyey:1941}
to model the angular distribution of aerosol-scattered light:
\begin{multline}
\betaa(589~\mathrm{nm},\theta,\htr)  \approx 4 \times 10^{-5} \cdot
\frac{1-g^2}{4\pi} \cdot 
\left(\frac{1}{(1+g^2-2g\cos\theta)^{3/2}} + f \frac{3 \cos^2\theta
  -1}{2\cdot (1+g^2)^{3/2}} \right) \cdot 
\exp(-\htr/\Haer)~\mathrm{m^{-1}} \quad.   \label{eq.betaa}
\end{multline}
Here, $g$ represents the mean value of
$\cos(\theta)$ 
and $f$ the strength of a second component to the backward scattering peak. 
Reference values of $g \approx (0.6\pm0.1), f \approx (0.4\pm 0.1)$ have been found by \citet{Louedec:2012} for a clear atmosphere and a desert-like environment in the Argentinean Andes.
%
%
Contrary to the Rayleigh scattering case on molecules, the value of $\alpha_{0,589~\mathrm{nm}}$ can
show large variations, depending on both the amount of aerosols and their composition. For
instance, a layer of Saharan dust (called ``calima'' on La~Palma) can dramatically increase the aerosol
scattering cross section~\citep{Lombardi:2008}. We neither take into acount these nor the possibility 
of clouds here, because their effects 
are considered more important obstacles for observation by
themselves than the scattered laser light.  Finally, we neglected any scattering contribution from stratospheric aerosols in the Junge layer\footnote{\re{The current stratospheric AOT amounts to 0.005, distributed over an altitude from 15 km to 30 km a.s.l. Residual scattered laser light from these altitudes get mostly focused into one camera pixel.}}.

\subsection{Computation of spurious LGS light on neighbour installation's pixels} 
\label{sec:geometry}
\begin{figure}
\centering
\includegraphics[width=0.4\linewidth]{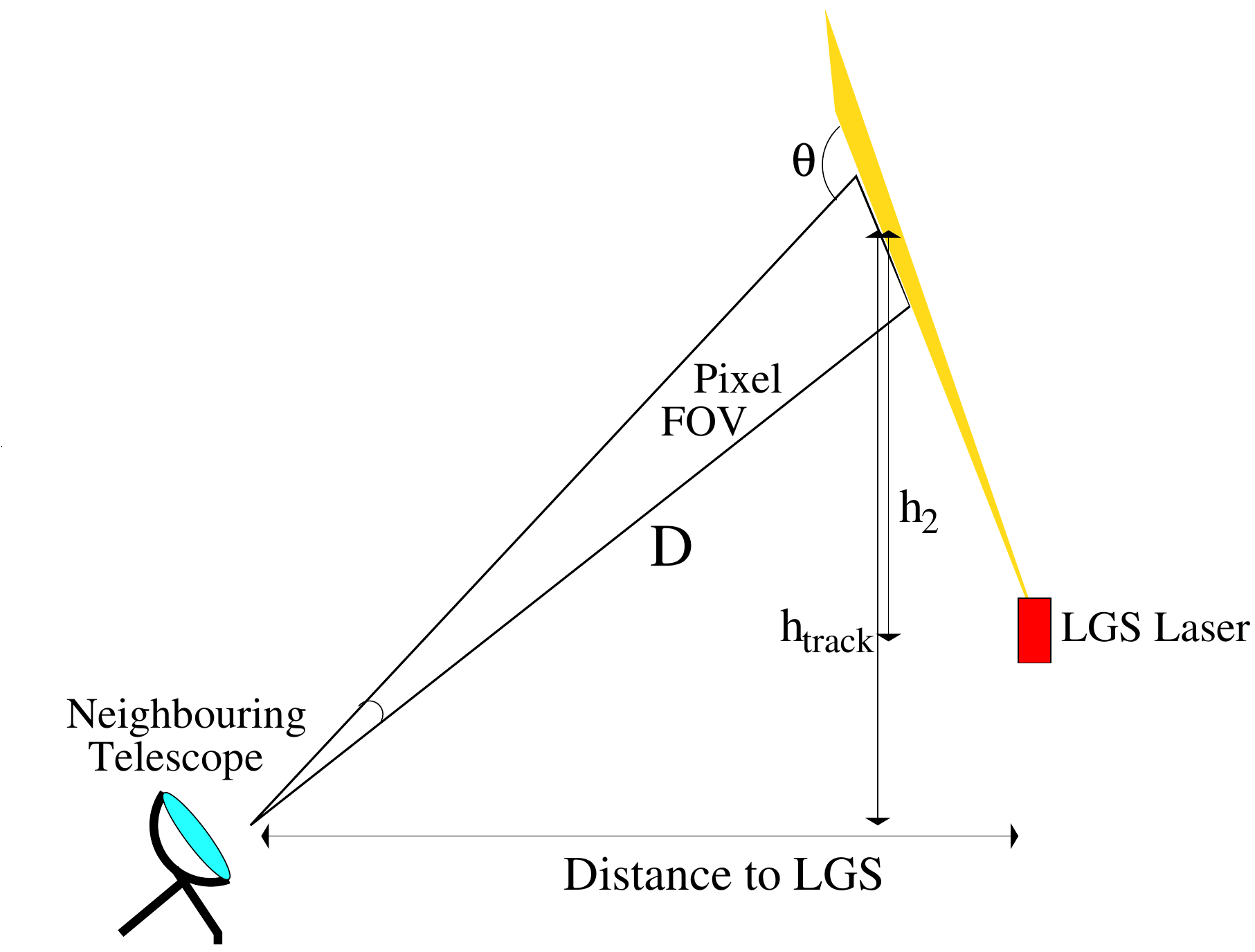} 
\caption{\label{fig:geometry} Sketch of the chosen geometrical
  conventions: The optical telescope points the LGS-laser in the direction of the neighbouring telescopes the laser light gets scattered 
  under an angle $\theta$ towards that telescope. The scattered light travels the distance $D$ from the scattering point to the CTA telescope mirror. 
  The track is then observed from an altitude $\htr$ with respect to the neighbouring telescope, and $h_2$ with respect to the LGS.}
\end{figure}

We assume that the neighbouring instrument 
collects light 
in a 
pixelated camera, 
(e.g., a CCD camera, or an array of Photomultiplier Tubes or silicon Photomultipliers). 
We compute the amount of light observed by a single pixel in the camera, and assume that the telescope observes the laser \re{uplink} beam under an angle $\theta$, 
such that $\theta=\pi$, if laser and the telescope's optical axis are
parallel, and $\theta=\pi/2$, if the axes cross perpendicularly,  
see \autoref{fig:geometry}.
The pixel observes a part of the laser track
$d_\mathrm{track}$ across its field-of-view ($\fovp$),  
at a distance $D$ from the laser \re{uplink} beam:

\begin{equation}
\dtrack = \frac{\fovp\cdot D}{\sin(\theta)} \cdot \re{ \Omega_\mathrm{blur}(D)} \quad,   \label{eq.track}
\end{equation}
\noindent
\re{where we have included the possibility that the laser beam width $w$ spreads over more than one pixel: }
\begin{equation}
  \Omega_\mathrm{blur} = 
  \left(1-\exp(-\frac{4r^2}{w^2})\right) \otimes \textit{PSF}_\perp(r/D) \quad.
\end{equation}
\noindent
\re{Here, $\textit{PSF}_\perp$ is the point-spread function of the telescope, projected onto the plane perperdicular to the laser propagation and $\delta$ the angular distance from the beam axis.}
We can, however, assume that the LGS laser is
sufficiently well collimated such that the observed beam width fits always into
one camera pixel \re{($\Omega_\mathrm{blur} \approx 1$)}, which is the case, at least, for the CTA cameras and their neighbouring LGS stations\footnote{%
Assuming a beam width of 0.4~m~\citep{li2016}, observed, in the absolutely worst case at 150~m (see \autoref{sec:scenarios}) distance by an MST, yields 2.6~mrad, smaller than the MST pixel size of 3~mrad,
or, at 300~m by an LST, yields 1.3~mrad, smaller than the LST pixel size of 1.7~mrad, \re{assuming that the optical aberrations are smaller than the size of a pixel, which is the case: Both MST and SSTs produce a point spread of about 0.02$^\circ$ (on-axis) to 0.05$^\circ$ towards the camera edges.}}. 

Assuming negligible loss of laser light due to scattering out of the beam, the observed photon rate inside one laser's track can be estimated from the total laser power, $P_\mathrm{laser}$  
(see \autoref{tab:laser_params}):
\begin{equation}
R_\mathrm{ph} = \frac{P_\mathrm{laser}\cdot \lambda_\mathrm{laser}}{h\,c} \approx \nlas \cdot 4.9\times 10^{19}\quad \mathrm{s^{-1}}~,
\end{equation}
where an individual laser power of 16.5~W has been assumed. The number of lasers simultaneously fired, $\nlas$, ranges from
one for the case of the GTC, to four in the case of VLT, to six in the case of the TMT and the
\re{ELT}. 

If the distance $D$ is large with respect to the camera dimensions
(which will always be the case during observations, at least for the CTA), the scattering angle can be approximated as 
constant for all pixels.
The light of the observed laser track scatters into a solid angle
$\Omega = \atel / D^2$, where $\atel$ is the telescope's mirror area. 
Considering a the mirror reflectivity $\xi$, \re{a transmission factor $T_\mathrm{optics}$ for the overall optics, filters and instrument}, and a photon detection efficiency $\pde$,
the photo-electron rate, which is then amplified by the dynodes of the photo-multiplier of the camera pixel, can be derived as\footnote{\autoref{eq.final} has been cross-checked with an independent LIDAR return power simulation program.} : 
%
%
\begin{equation}
\rpix = R_\mathrm{ph} \cdot \tair(\theta,h) \cdot \xi \cdot T_\mathrm{optics} \cdot  \pde \cdot \beta(\theta,\htr) \cdot \frac{\atel}{D^2} \cdot \frac{\fovp\cdot D}{\sin\theta}  \quad,
\label{eq.final}
\end{equation}
where
\begin{align}
\beta(\theta,\htr) &= \left(\betam(589~\mathrm{nm},\theta,\htr) +\betaa(589~\mathrm{nm},\theta,\htr) \right)
\nonumber \\
 & \approx  \quad \bigg(\frac{ 0.95 \cdot \cos^2\theta + 1 }{2} \cdot e^{-(\hcta+\htr)/\Hmol}
 +  \re{1.5} \cdot \left( \frac{1}{(1.13-\cos\theta)^{3/2}} + 0.17\cdot(3\cos^2\theta-1)   \right) \cdot e^{-\htr/\Haer} \bigg) \nonumber\\
 & \qquad   \cdot 10^{-6}~\mathrm{m^{-1}} \quad,
\end{align}
\noindent
as derived in Eqs.~\ref{eq.betam} and~\ref{eq.betaa}, with the average reference values for $g$ and $f$ inserted. 
Moreover, we have included an atmospheric transmission factor, $\tair$, from the laser light dispersion point to the telescope mirror.

\bigskip
Besides the rather obvious observation that those cameras will be affected most which show the highest combination of the factors $T_\mathrm{optics} \cdot \fovp\cdot \atel \cdot \pde / D$, \autoref{eq.final} requires the following comments:
\begin{enumerate}
\item The distance $D$ to
  the laser beam reduces the amount of registered light in a \textit{linear}
  way. This is due to the combination of reduced solid angle (which
  scales with $D^{-2}$) and the increased part of the track spanned by
  the FOV of a pixel (which scales with $D$, due to the one-dimensional
  propagation of the laser beam). 
\item The function $(1+\cos^2(\theta))/\sin(\theta)$ has a
  (divergent) maximum at $\theta=\pi$, i.e. when the beam propagates along the telescope's optical axis.
\autoref{eq.track} assumes then that the pixel integrates the 
  light beam extending to infinite. In the case of such large scattering angles, the development of the scattering coefficient \textit{across the FOV of the pixel} needs
  to be taken into account, and \autoref{eq.track} translates into the geometrical overlap function of the LIDAR equation~\citep{fernald1972}.
  As a matter of fact, $(1+\cos^2(\theta))/\sin(\theta)$ increases only by a
 factor four at $\theta =  0.85\pi$ with respect to its minimum at $\pi/2$, which suggests that \autoref{eq.final} 
 is at least not valid for viewing angles $\theta \gtrsim  0.85\pi$\footnote{%
   \re{Note that such high scattering angles are actually possible if the laser propagates away from the neighboring installation at maximum zenith angle.
   In that case, $\theta_\mathrm{max} \approx \pi - L/\htr \cdot \cos^2\thetal$. However, in such cases, the scattered return flux of light is negligible, as we will see later.}  }.
\end{enumerate}

\ree{In case of observation of a same source by both the LGS-equipped and the neighbouring telescope, the photo-electron rate received by the outmost camera pixel of the neighbouring telescope can be approximated as:}
  \begin{equation}
    \rpix \approx R_\mathrm{ph} \cdot \tair(\theta,\htr) \cdot \xi \cdot T_\mathrm{optics} \cdot  \pde \cdot \beta(180^\circ,\htr) \cdot \frac{\atel \cdot \fovp}{L} ~, \label{eq:parallel}
  \end{equation}
\noindent
\ree{where $L$ is the distance of the neighbouring telescope to the AO laser system (see Table~\ref{tab:laser_params}), $\thetal$ the zenith angle of the AO-laser and $\htr \approx 2L\cdot \cos\thetal/\fovc$.}

\re{Additionally to Rayleigh and Mie scattering, the telescope may observe fluorescence emission from the excited sodium layer itself.}
\re{Assuming an average coupling efficiency of $(140 \div 160)$~m$^{2}$~s$^{-1}$~W$^{-1}$ of circularly polarized laser light at 589.159~nm to sodium atoms~\citep{Jin:2014a,Lu:2016} and a vertical column density of sodium of $(3 \div 6)\times 10^{13}$~m$^{-2}$~\citep{Moussaoui:2010},
  we can derive an average effective volume scattering coefficient of:}
\begin{equation}
  \beta_\mathrm{Na}(\theta) \approx (1.7 \pm 0.6) \times 10^{-7} \cdot \left( 1 + \cos^2\theta\right) \cdot \left(2.25 - 1.25\cdot \sin\alpha\right) ~\mathrm{m}^{-1} ~, \label{eq.betana}
\end{equation}
\noindent
\re{where $\alpha$ denotes the angle between the laser beam propagation and the direction of the Earth's magnetic field lines.
  The last factor is valid if $\sim 10$\% of the laser light is used to simultaneously excite the $F=1$ hyperfine ground state of sodium with $589.157$~nm wavelength (``optical pumping'')~\citep{Moussaoui:2008}}.
\re{The scattering angle dependency stems from the polarization of the laser light~\citep{Steck:2010}.}
\re{Eq.~\ref{eq.betana} assumes a constant scattering efficiency throughout the layer, which shows in reality complicated structures~\citep{Neichel:2013}.}
\re{Nevertheless, we will use it, together with Eq.~\ref{eq.final}, to roughly estimate the photo-electron rate received from the illuminated sodium layer. Our final results do not depend on the fine-structure of that layer.} 

\bigskip
The concrete case of the number of spurious LGS light on CTA camera pixels will be discussed in Sec.~\ref{sec:scenarios}.

\clearpage
\section{Probability of beam crossings}
\label{sec.prob}
In this section, a quantitative estimation of the ``collision'' probability 
  between the LGS system and nearby telescopes is computed, 
  together with the amount of observation time disturbed by the LGS, or even lost, for a neighbouring installation.
  As above, results are discussed and computed for the case of the
CTA, but the used formulae are generic and can be adapted to different
facilities.

\bigskip
We assume that an LGS impedes observation of a certain strip in
the sky, if the photon rate in a series of pixels received by
at least one neighbouring instrument's camera becomes larger
than a maximally acceptable critical threshold $\rcrit$.
We first notice that if a Laser Traffic Control System (LTCS)~\citep{Summers:2003} is used in its \textit{basic configuration}\footnote{Basic configuration means here using a strict ``first-on-target'' (or ``first-come-first-serve'') policy.}~\citep[see, e.g.,][]{Summers:2012}, \re{currently used for the ORM and the Paranal Observatory},
the affected sky region can be technically avoided.
Therefore, for most of the steady sources, the scheduling system can take into account the LTCS information and
re-schedule a source to later times if necessary.  
The situation is different in case of targets that are either $a)$ part of
multi-wavelength or multi-instrument campaigns (and therefore
observed with pre-defined, fixed observation times) or $b)$ that are the
result of fast Target of Opportunity (ToO) alerts.
The former are scheduled well in advance in coordination with other facilities and their scientific merit relies on contemporaneous data taking, while the latter are observations motivated by external triggers
or other activators demanding immediate reaction and repositioning of the telescopes (such as, e.g., gamma ray burst or gravitational wave alerts or generally, flaring sources). 
In these cases, adequate scheduling of sources is practically impossible without a high risk of losing the science case. 
The frequency of such ToO alerts,
the duration of their follow-up observations, and the further characteristics of the campaigns depend on the specific science case and the specificities of the neighbouring facility.
Nevertheless, a general computation of the interference probability is hereafter attempted.

\begin{figure}
\centering
\includegraphics[width=0.45\linewidth]{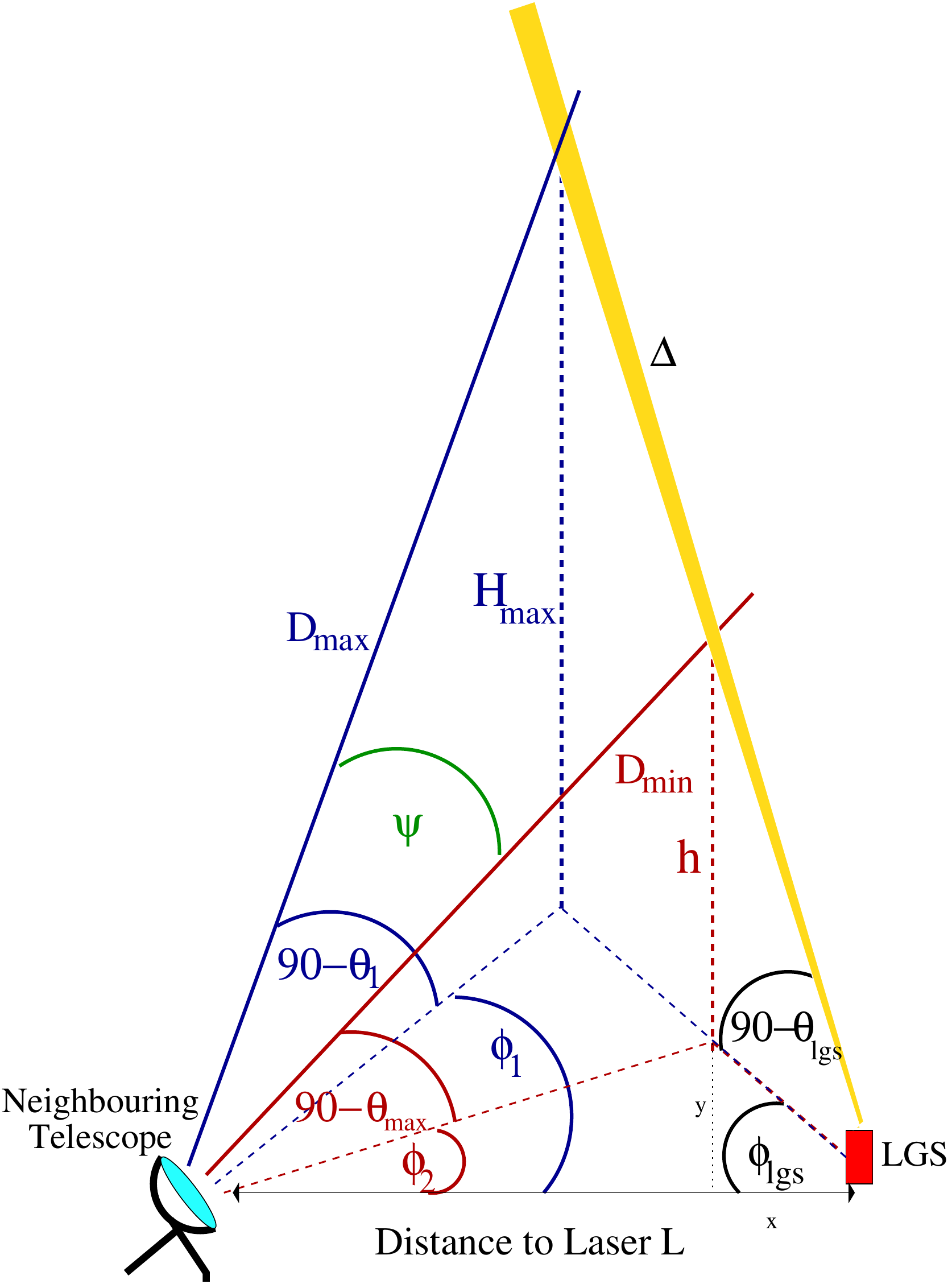}
\caption{\label{fig.geometry} Sketch of the geometry to define the angle of the observability cone vetoed by the LGS. }
\end{figure}

\bigskip
We start by defining the geometry of the problem in \autoref{fig.geometry}: an LGS is located at a horizontal distance $L$ of an affected telescope;
the LGS beam vetoes a band across the sky whose width can be assumed to be the (larger) FOV of that neighbouring instrument.  This band may cross the observability cone of the close-by telescope,
i.e., the region of the sky accessible by the telescope,
which itself extends from zenith to the largest observable   zenith angle $\thetam$, spanning $2\pi$ in azimuth for extra-galactic targets,
or otherwise, the galactic plane, until reaching $\thetam$.
The vetoed band starts at the point at which the laser is seen under a
zenith angle $\thetam$ by the neighbouring instrument, at an altitude $h$ and a distance $\Dmin$.
The band ends when the laser reaches a maximum altitude $\Hm$ above which
the laser-induced spurious photon rates in the neighbouring instrument's cameras ($\rpix$ from \autoref{eq.final}) falls below $\rcrit$.
At this point, the laser beam has a distance $\Dmax$ from the neighbouring telescope. 
The vetoed band has a length $\Delta$ and is seen from the neighbouring instrument under an angle $\Psi$. 
Using the estimation for $\rpix$, \autoref{eq.final}, we can derive $\Hm$ (for details, see appendices \ref{app:distance} through \ref{app:Hmax}).

\bigskip
The condition that the LGS light enters the neighbouring instrument's observability cone at all is computed hereafter. 
First, we define the projected distances of the laser beam on ground: 
\begin{align}
   \psi_x  &= \tan\thetal \cdot \cos\phil  \\
   \psi_y  &= \tan\thetal \cdot |\sin\phil| \quad,  
\end{align}
\noindent
where $\phil$ has been defined such that the direction towards the neighbouring instrument
defines $\phil =0$. If multiplied with an altitude $h$, both yield the corresponding distances $(x,y)$, shown in \autoref{fig.geometry}.
The condition $\Theta(\thetal,\phil)$, that the laser light enters at all the neighbouring instrument's observability cone is then given by (see \autoref{sec.derivobs} for details): 

\begin{equation} \label{eq.entering} 
 \Theta(\thetal,\phil): \quad \left\{
   \begin{array}{cc}
      \psi_y < \tan\thetam   &  \textrm{for~$\Hm \cdot \psi_x \geq L$}  \\[0.2cm]
       \tan^2\thetal - 2\ddfrac{L}{\Hm}\psi_x + \left(\ddfrac{L}{\Hm}\right)^2 {} < {} \tan^2\thetam & \textrm{otherwise}~.
   \end{array}  \right. 
\end{equation}

\begin{figure}
\centering
\includegraphics[width=0.24\linewidth]{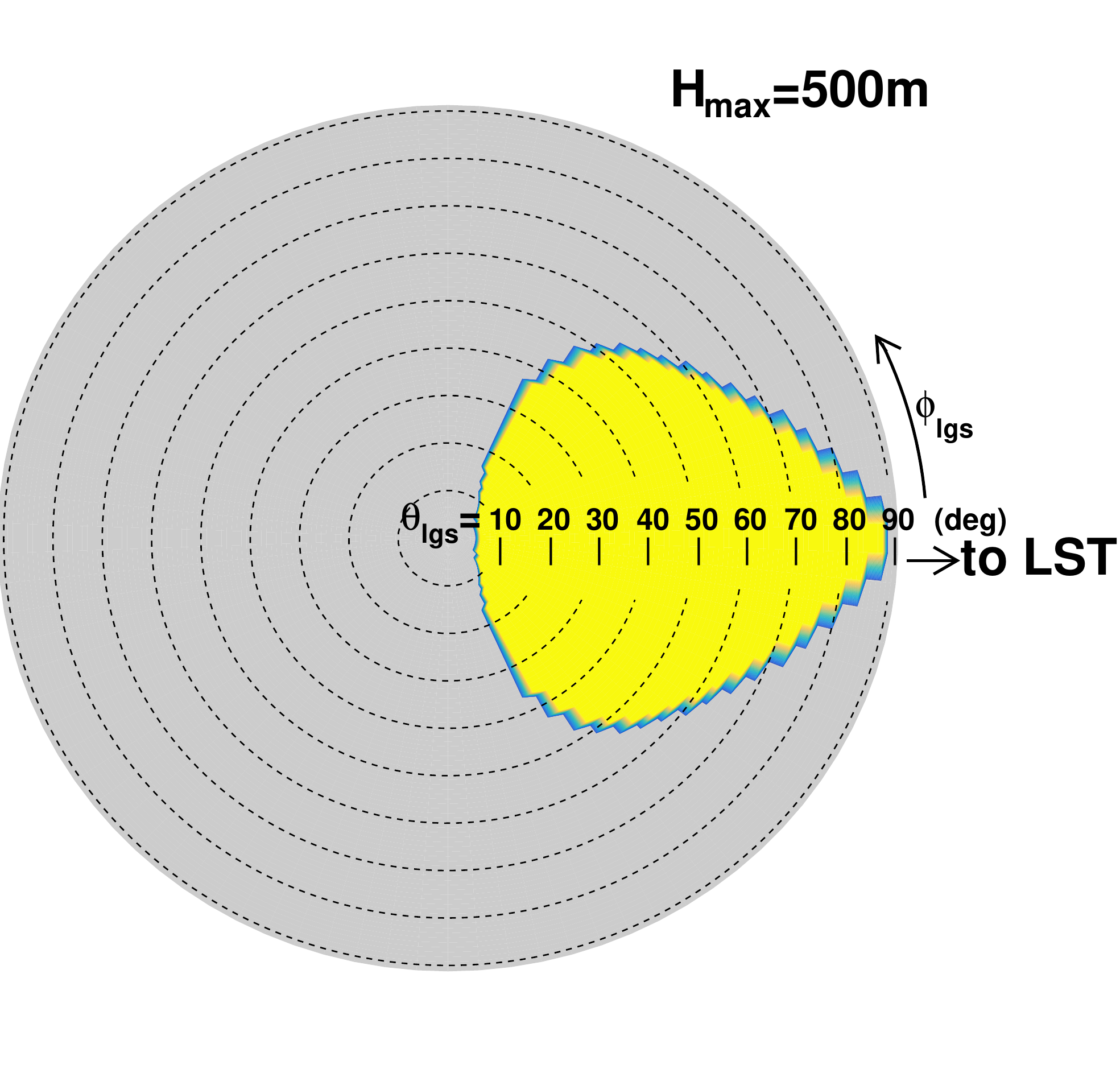}
\includegraphics[width=0.24\linewidth]{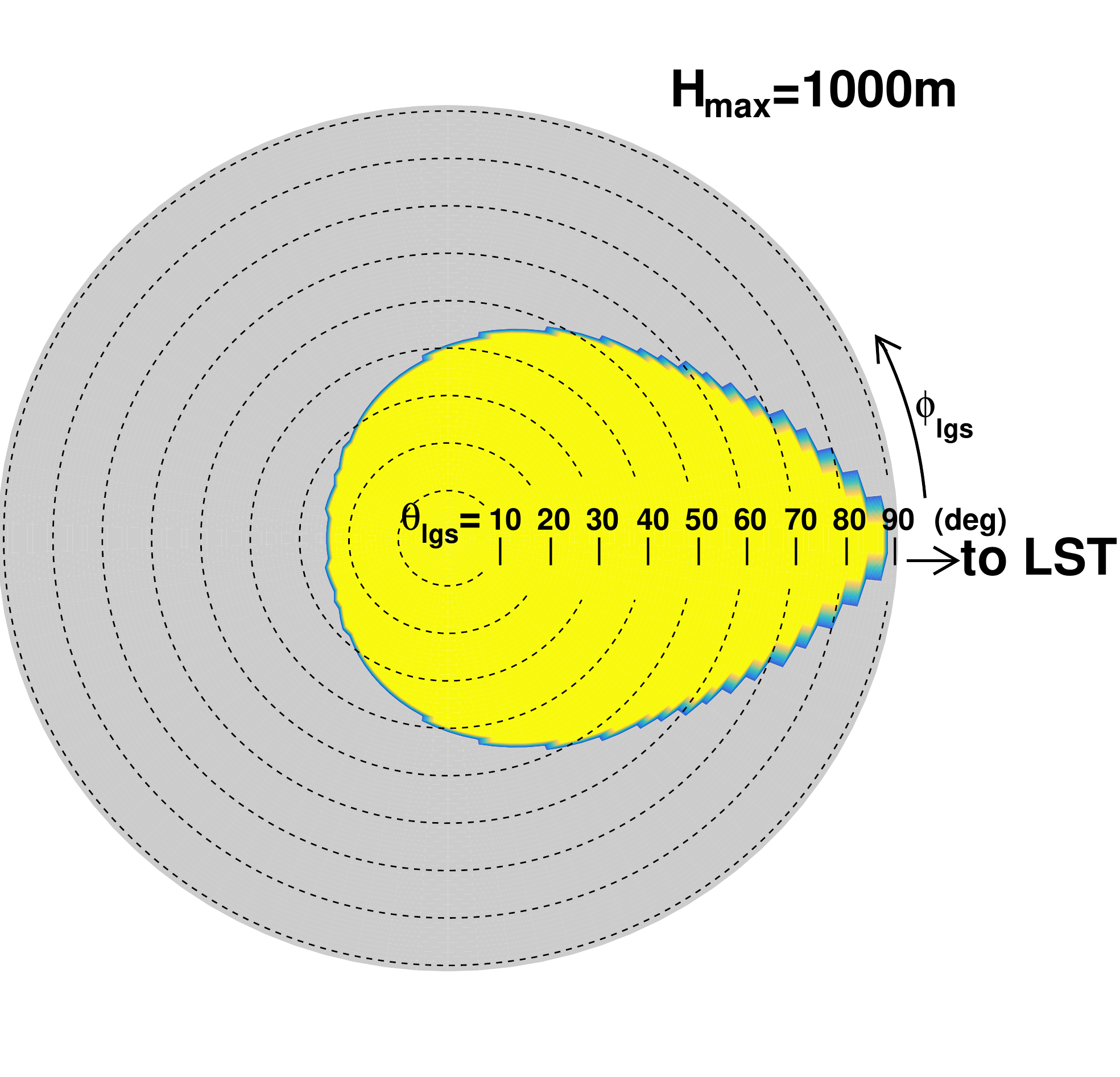}
\includegraphics[width=0.24\linewidth]{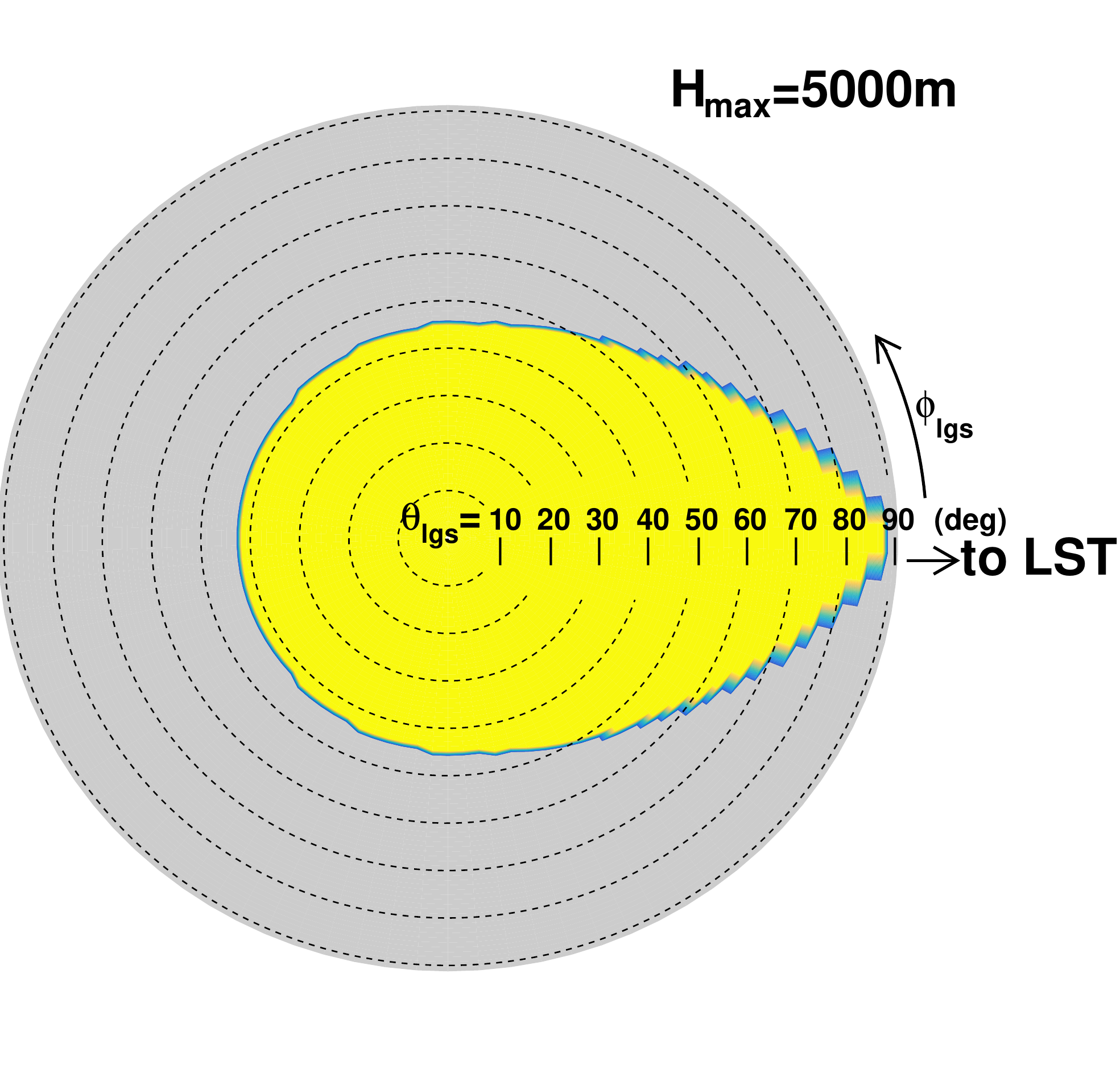}
\includegraphics[width=0.24\linewidth]{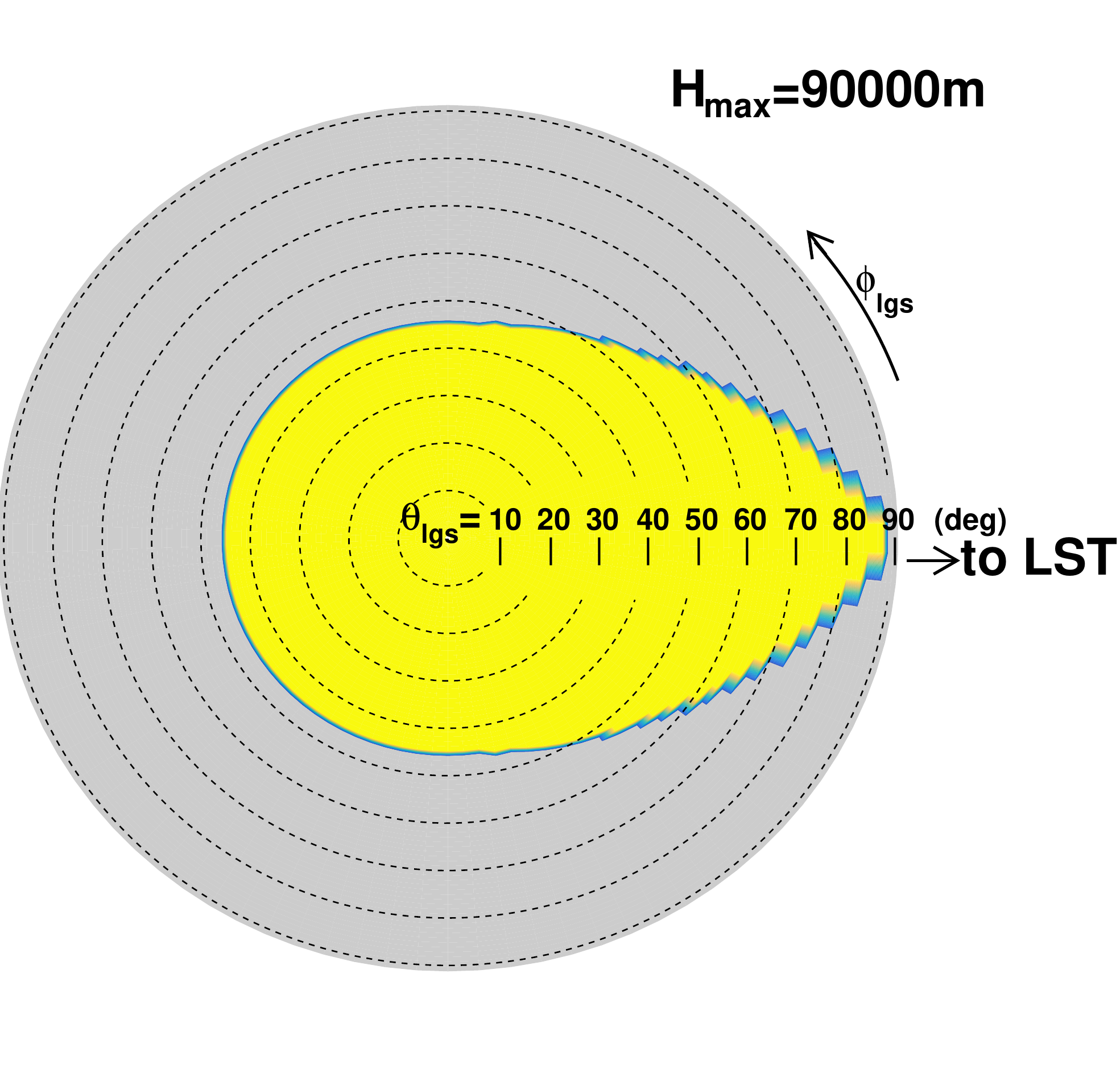}
\caption{\label{fig.obscondition} LGS pointings $\thetal,\phil$ fulfilling condition Eq.~\protect\ref{eq.entering},  
  i.e. to enter at all the observability cone (in this case opened by a CTA-N LST), for four arbitrary values of $\Hm$. 
  The laser is located in the center here, the direction to the LST telescope pointing towards its right side. 
  A maximum zenith angle of \re{$45^\circ$ has been adopted for the CTA telescope and}
  $90^\circ$ for the AO-laser, to highlight the structure of the beam crossing regions. }
\end{figure}

Depending on the zenith and azimuth angles ($\thetal,\phil$)  of the
actual LGS pointing, the vetoed band may be larger or shorter, or even
null (see \autoref{fig.obscondition}, where  one can see  that not all LGS pointings will be able to
generate a conflict (``collision'') with the neighbouring telescope, particularly if
they point away from it).

For those LGS pointing directions, which fulfill condition \autoref{eq.entering}, we can calculate the altitude $h$, at which the laser   
beam enters the observability cone of the neighbouring telescopes. After solving several
geometrical relations (see again \autoref{sec.derivobs}), we obtain:
\begin{equation}\label{eq.exclusion}
 h = \frac{L}{\psi_x + \sqrt{\tan^2\thetam - \psi_y^2}}\quad . 
\end{equation}

\begin{figure}
\centering
\includegraphics[width=0.45\linewidth]{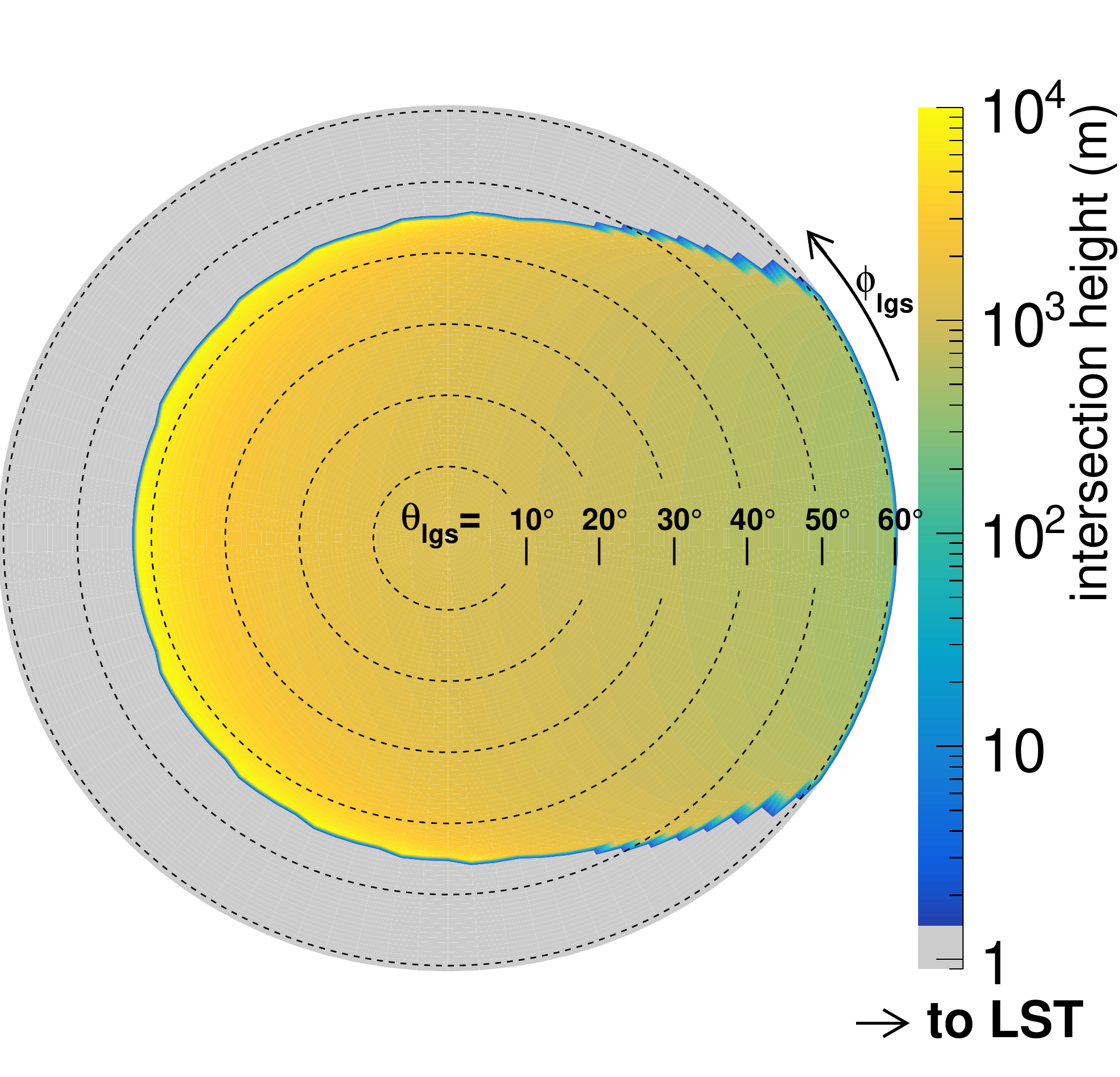}
\caption{Polar plot of the intersection height (in meters) of the six LGS lasers of the TMT with the observability cone of an LST for fast ToO alerts in spherical coordinates $\thetal,\phil$.  
Here, a maximum zenith angle of $60^\circ$ has been adopted for the LGS. The laser is located in the center, the direction to the LST points towards its right side.
Note the logarithmic scale of the color palette. \label{fig.observability}} 
\end{figure}

\autoref{eq.exclusion} defines the intersection height of the laser with the observability cone of the neighbouring instrument, and is shown as an example in \autoref{fig.observability}, for 
the case of six TMT lasers intersecting with the observability cone of a CTA-N LST. \\

The angular length $\Psi$ of the laser
beam, as seen from the location of a telescope within the observability cone, from zenith to $\thetam$, is then:

\begin{eqnarray}\label{eq.cospsi}
        \cos(\Psi) &=& \frac{1}{2} \cdot \Bigg\{  \left( \frac{\Hm}{h}\right)^2 \cdot \frac{\cos\thetam}{\cos\theta_1} + \left(\frac{h}{\Hm}\right)^2\cdot \frac{\cos\theta_1}{\cos\thetam}  -  \nonumber\\
               &&      - \frac{\cos\theta_1\cos\thetam}{\cos^2\thetal} \cdot \left(1+  \left(\frac{h}{\Hm}\right)^2 - 2\frac{h}{\Hm} \right) \Bigg\}     \\[0.2cm]
     && \mathrm{with:} \nonumber\\[0.2cm]
        \frac{1}{\cos\theta_1} &=& \frac{1}{\cos\thetal} \cdot \sqrt{ \left(\frac{L\cos\thetal}{\Hm}\right)^2 - 2 \frac{L\sin\thetal\cos\thetal}{\Hm}\cos\phil + 1 } ~,
\end{eqnarray}

and $h$ as defined in \autoref{eq.exclusion}. An explicit version of \autoref{eq.cospsi}, with all values inserted, is derived in the appendix \ref{app.psi}. \autoref{fig.angularlength} displays $\Psi$ as a function of the LGS pointing angles,
again for the case of six TMT lasers shining into the observability cone opened by a CTA-N LST.

\begin{figure}
\centering
\includegraphics[width=0.45\linewidth]{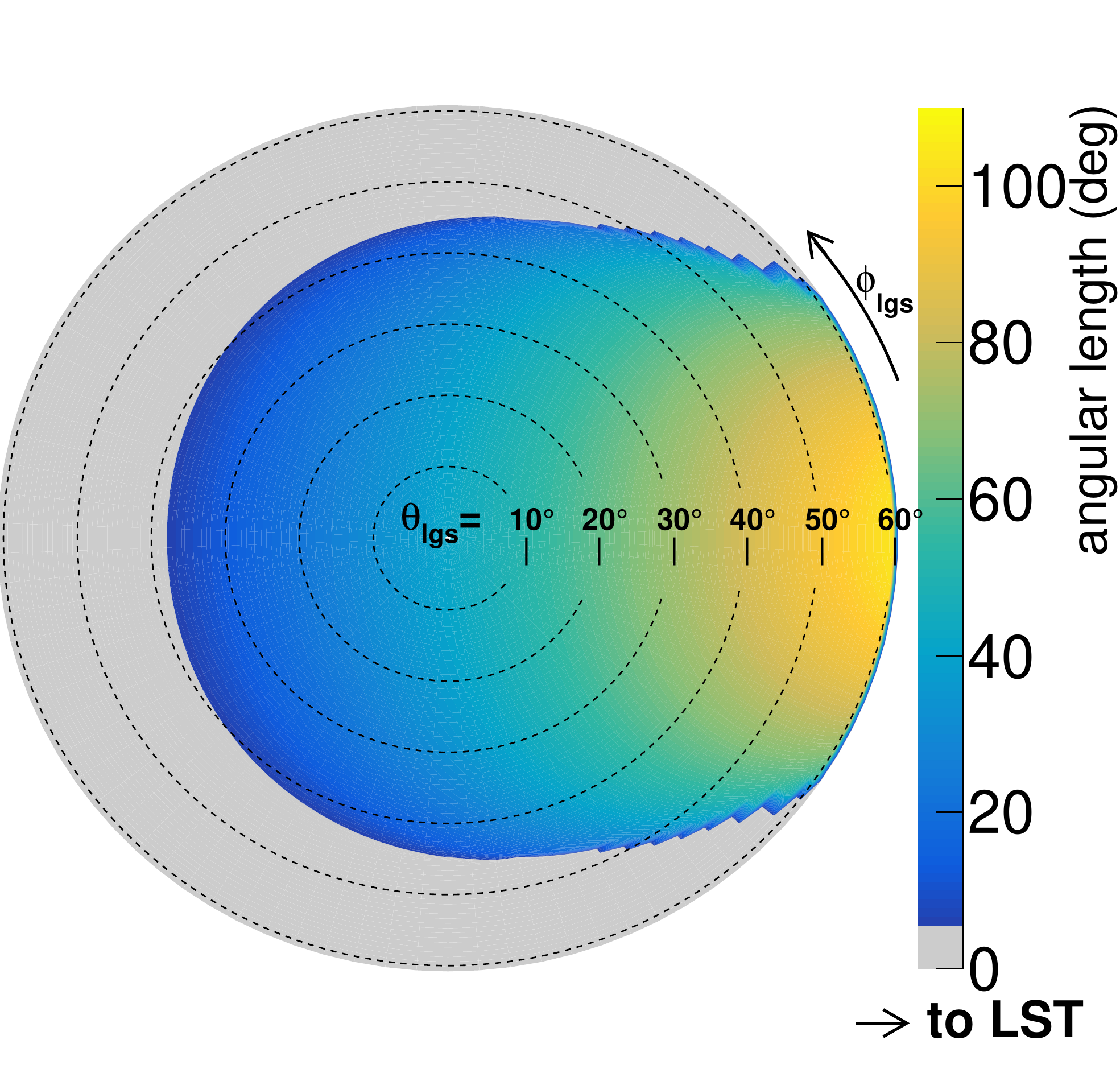}
\caption{Angular distances $\Psi$ (in degrees) covered by the LGS beam in polar coordinates ($\thetal,\phil$), 
as seen from the CTA-N telescopes within its observability cone for fast ToO's (opened from $\thetam=0$ to $\thetam=45^\circ$).
 The laser is located in the center here,  the direction to CTA-N points towards its right.
A maximum possible zenith angle of $\thetal = 60^\circ$ has been adopted.
\label{fig.angularlength}} 
\end{figure}

The part of the observable sky, which is found vetoed by an LGS laser, can then be modeled as: 

\begin{equation}\label{eq:pveto}
\pveto(\thetal,\phil) = \frac{\alpha(\thetal,\phil) \cdot \textit{FOV}_\mathrm{vetoed}}{\Omega_\mathrm{obs}} \quad, 
\end{equation}

\noindent where the total solid angle  $\Omega_\mathrm{obs}$, available for observations by the neighbouring telescope can be computed as: 
\begin{itemize}
\item $\Omega_\mathrm{obs,extra-gal} =  2\pi \cdot (1-\cos\thetam)$ for extra-galactic targets of the neighbouring telescope.
\item $\Omega_\mathrm{obs,gal}       = |b_\mathrm{gal}| \cdot 2 \thetam  \sim 0.3~\mathrm{Sr}$ for galactic observations of both the neighbouring telescope.
We assume an average diameter of the observable Milky Way $|b_\mathrm{gal}| \sim 0.2$~rad and that the Milky Way passes through very close to zenith\footnote{%
This is a reasonable assumption for the required case when both LGS and the neighbouring telescope observe a Galactic source.}. 
\end{itemize}

For the vetoed observation band width, we use an effective telescope field-of-view $\textit{FOV}_\mathrm{tel}$, considered larger than the laser beam width,
and a vetoed observation band length:
\begin{itemize}
\item $\alpha = \Psi(\thetal,\phil)$ for extra-galactic LGS pointing targets.
\item $\alpha = \Theta(\thetal,\phil) \cdot b_\mathrm{gal} / \avg{\sin\delta}$ for galactic pointings of both the LGS and the neighbouring telescope, and $\avg{\sin\delta} \sim \sin(\pi/4)$ to account for the average tilt $\delta$ of the Milky May with the LGS beam.
The condition $\Theta(\thetal,\phil)$ that the laser enters the observability cone at all, is taken from \autoref{eq.entering}.
\end{itemize}

\begin{figure}
\centering
\includegraphics[width=0.45\linewidth]{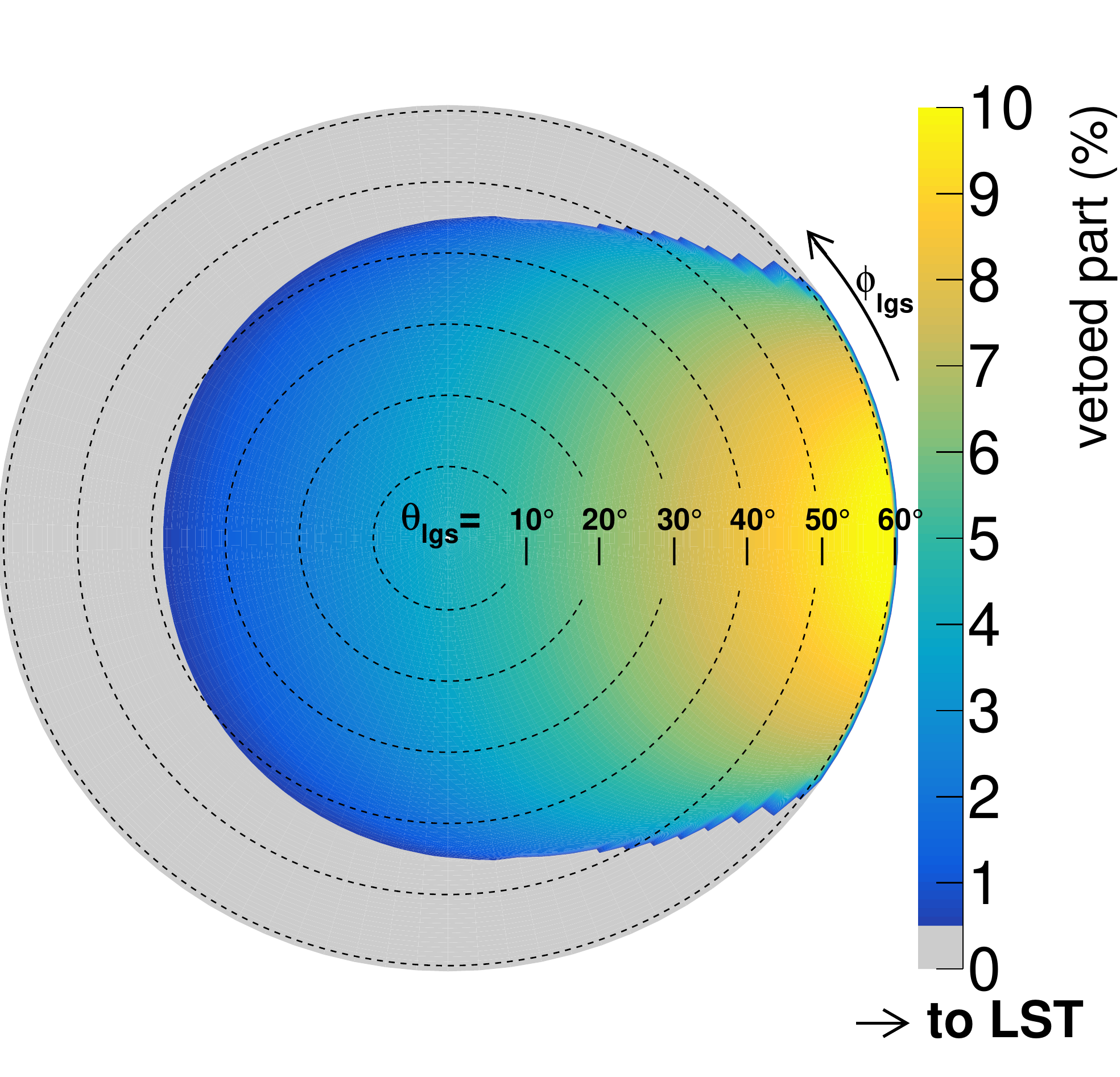}
\caption{Part of the observability region $\pveto$ (shown here in as percentage) for extra-galactic CTA-N ToO's vetoed by the LGS beam, depending on its pointing angle in 
spherical coordinates. Angular conventions as in the previous figures. \label{fig.percentagevetoed}} 
\end{figure}

\autoref{fig.percentagevetoed} shows $\pveto$ for the extra-galactic case and six TMT lasers shining into the observability cone of an LST. 

All previous formulae have been derived in a coordinate system where the direction from the LGS to the neighbouring telescope defines $\phi = 0$.
We want to estimate the probabilities of LGS pointings in local coordinates $(\thetal,\phi_\mathrm{lgs,orig})$, defined by $\phi_\mathrm{lgs,orig} = 0$ when the LGS points to the North,
and rotate one coordinate system to the other:
\begin{equation}
(\thetal,\phil)=(\thetal,\phi_\mathrm{lgs,orig}-\delta_\mathrm{tel}) \quad,
\end{equation}
\noindent
where $\delta_\mathrm{tel}$ is the angle between the line connecting
the neighbour telescopes and the LGS and the North-South axis. 

\begin{figure}
\centering
\includegraphics[width=0.49\linewidth]{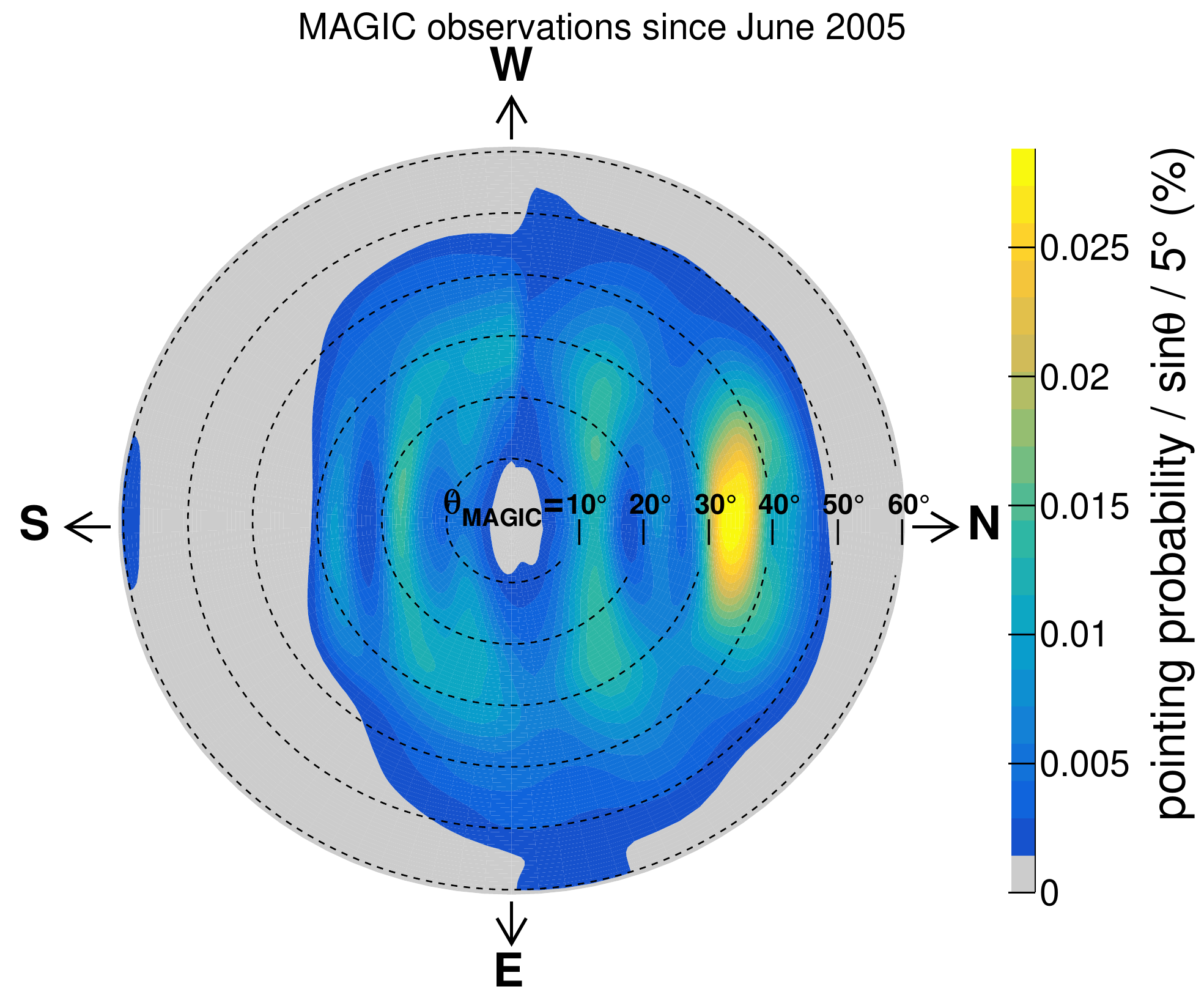}
\includegraphics[width=0.49\linewidth]{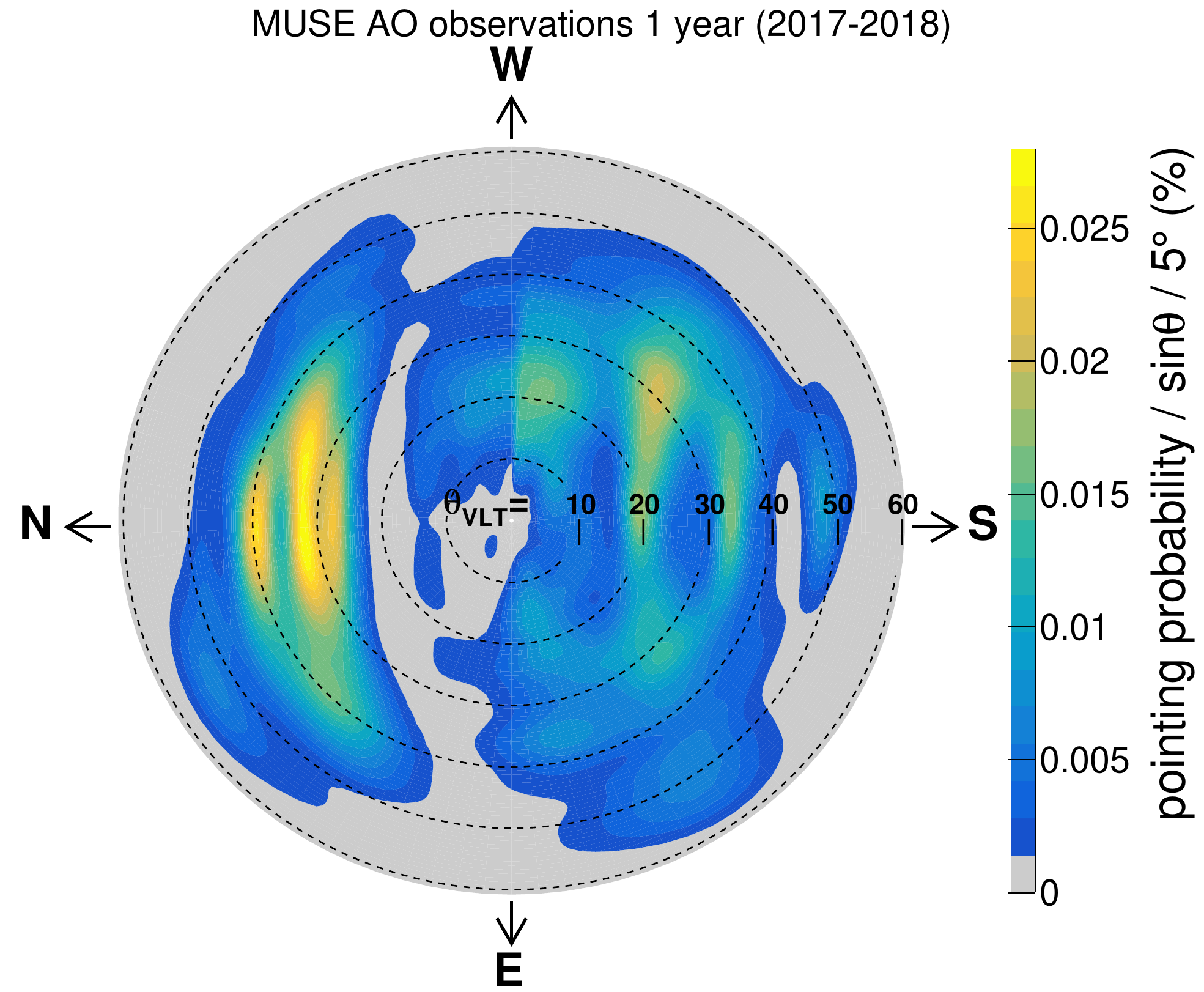}
\caption{\label{fig:pointings} 10 years of MAGIC pointing history in local spherical coordinates ($\theta,\phi$), smoothed and corrected for the most frequently observed calibration source
\re{and one year of VLT pointing history with the MUSE instrument in Wide-Field-Mode, using adaptive optics}. 
%
}
\end{figure}

Further, a probability distribution function of LGS pointings is needed: 
Since this is not possible to do, before an actual observation schedule is produced, we make a best guess using 10~years of the MAGIC telescopes'~\citep{Aleksic:2014lkm} pointing history\footnote{%
MAGIC is a currently operating instrument of the same class as CTA, located at the same site.}
For comparison, we also checked the local pointing field of one year of GTC pointings (courtesy of Antonio Luis Cabrera Lavers) \re{and a bit less than one year of MUSE\footnote{\url{https://www.eso.org/sci/facilities/paranal/instruments/muse/inst.html}} Wide-Field-Mode observations carried out with adaptive optics.}  
\re{We find compatible results, if the different proportions of galactic and extra-galactic targets are taken into account}.
\re{Varying the different pointing probability maps, the final results shown in \autoref{fig:pointings} differ by less than 20\%.}
Since the MAGIC observations were
  dominated by few reference and calibration sources, we smoothed
  the histogram using a kernel algorithm acting on a $5\times 5$
  cell~\citep{rootsmooth}.  The outcome is shown in \autoref{fig:pointings}. 

The pointing probability map of \autoref{fig:pointings} needs to be rotated from $(\thetal,\phi_\mathrm{lgs,orig})$ to the local coordinate frame $(\thetal,\phil)$ and convoluted with the probability for laser beam vetoes:

\begin{equation}\label{eq.pconflict}
P_\mathrm{conflict}(\thetal,\phil) = \ddfrac{\pveto(\thetal,\phil) \cdot \pobs(\thetal,\phil)}
                                      {\int_0^{2\pi} \int_0^{\theta^\mathrm{max}_\mathrm{lgs}}\!\!\! \pobs(\thetal,\phil) ~\sin\thetal~\mathrm{d}\thetal \mathrm{d}\phil}  ~,
\end{equation}

The map $P_\mathrm{conflict}(\thetal,\phil)$ yield the differential probability to reside in a certain pointing direction times the 
probability to veto the telescope pointings in that direction. Examples for the CTA-N are shown in \autoref{sec:results_prob}.

\bigskip
Finally, we compute the total probability for a neighbouring telescope observation vetoed by an LGS laser by integrating \autoref{eq.pconflict} and multiplying with the duty cycle $\eta$ of the LGS system:
\begin{equation}\label{eq.pcollision}
P_\mathrm{conflict} = \eta \cdot \int_0^{2\pi} \int_0^{\theta^\mathrm{max}_\mathrm{lgs}}\!\!\! P_\mathrm{conflict}(\thetal,\phil) ~\sin\thetal~\mathrm{d}\thetal \mathrm{d}\phil ~.
\end{equation}
\noindent
Here, the duty cycle of the LGS is assumed to be constant over the entire zenith angle range up to the maximum zenith angle foreseen for observations with the LGS, $\theta^\mathrm{max}_\mathrm{lgs}$.

\bigskip
We will apply the above formulae to a realistic case, the one of the CTA-N, in \autoref{sec:results_prob}.


\section{Results}
\label{sec:results}

In this section, we make use of the general formulae computed in the
previous section and apply them specifically to the case of the CTA. 
In \autoref{sec:scenarios}, we compute the amount of LGS 
light scattered onto CTA camera pixels, using the formalism of \autoref{sec:computation}, and in \autoref{sec:results_prob}, we compute the
probability of interference with the LGS lasers under best-guessed
observation conditions, using the formalism of \autoref{sec.prob}.

\subsection{LGS induced light on CTA pixels}
\label{sec:scenarios}

%


We insert a mirror reflectivity of \re{$\xi\approx 0.85$} (assuming SiO$_2$
and HfO$_2$ coated aluminum mirrors~\citep{pareschimirrors}), \re{a camera protection window transparency of 0.92 for $T_\mathrm{optics}$}, an
altitude of $h_\mathrm{CTA} \approx 2200$~m for both sites, and
  atmospheric transmission for the air, \re{ranging from $\tair = 0.93^2 \approx 0.86$ to $0.97^2 \approx 0.94$}\,\footnote{%
   From~\citet[Fig.~3 of][]{patat2011}, we obtain about 0.025~mag/airmass aerosol extinction and from Fig.~1 about 0.95 for the total molecular transmission from ground to infinity,
   for vertical incidence. Scaling to about 10~km above ground, this translates to \re{0.94} for the overall transmission at 589.2~nm. For very close scattering, molecular transmission
   may be negligible, yielding only $\tair \approx 0.97$ from aerosol transmission, while larger inclination angles (hence airmass) may lead to transmissions down to \re{0.91} for extreme cases.
   Hence, for clear nights, the transmission estimate may change by up to 2\%, depending on the observation angle and the interaction altitude of the laser light.}
  \re{for the scattered laser light}\footnote{%
     for the light emitted by sodium fluorescence, we use $\tair \approx 0.85^2 = 0.72$.} into \autoref{eq.final} and obtain for the
LGS-induced photo-electron rate onto a CTA camera pixel:  
%
%
\begin{eqnarray}\label{eq.final2}
  \rpix & \simeq & (\re{5.2} \cdot 10^{13}~\mathrm{m}^{-1}) \cdot \nlas \cdot  \pde \cdot
  \frac{\fovp\cdot \atel}{D\cdot \sin(\theta)} \cdot \bigg(\re{0.26} \cdot
  ( 0.95\cos^2\theta + 1 ) \cdot e^{-\htr/\Hmol} \nonumber \\
  & &  +   \left( \frac{1}{(1.13-\cos\theta)^{3/2}} + 0.17\cdot(3\cos^2\theta-1)   \right) \cdot e^{-\htr/\Haer} \bigg)~\mathrm{s^{-1}}\quad.
\end{eqnarray}
\noindent
Here, 
$\pde$ denotes the photon detection efficiency (PDE) of the
 photon detector at 589.2~nm, and the relative distances between the CTA telescopes and the LGS, $D$, are reported in \autoref{tab:laser_params}.
 
The CTA telescopes (LST, MST, SST) dish sizes and camera pixel
fields-of-view, as found in~\citet{Actis:2011}, are reported in
\autoref{tab2}\footnote{%
  see also \url{https://www.cta-observatory.org/project/technology/}}.
Until very recently, photomultiplier tubes (PMTs) have
been the common choice for equipping IACT cameras, due to their large
photon detection efficiency (PDE) from 300~to 450~nm, large size and
fast time response. However, silicon photomultipliers (SiPMs) are
emerging as an interesting alternative. This rapidly evolving
technology has the potential to become superior to that of 
PMTs in terms of PDE, which would further improve the sensitivity
of IACTs, and provide a price reduction per
detector area. An example of a working SiPM-based IACT is
FACT~\citep{Anderhub:2013cqa}. In CTA, this choice is already the
default for the double-mirror Schwarzschild-Couder SSTs (labelled SST-SC
in \autoref{tab2}) as well as the single-mirror Davies-Cotton SST
(labelled SST-DC in \autoref{tab2}). Schwarzschild-Couder optics
demand a compact camera~\citep{Bonanno:2016}. However, the SiPM choice was also adopted
for the SST-DC that makes use of the non-commercial Hamamatsu
S10943-2832(X)~\citep{Heller:2016rlc}. The MST
telescopes will instead host PMT-based
cameras~\citep{Glicenstein:2016dzr}, using the R12992-100 PMT from
Hamamatsu\,\footnote{\url{http://www.hamamatsu.com/us/en/R12992-100.html}},
and for the proposed Schwarzschild-Couder MST~\citep{2014SPIE.9145E..33M} at CTA-S.
The case of LST is peculiar: its baseline design will make use of the PMT
Hamamatsu R19200-100~\citep{Okumura:2015}, with the four LSTs planned
for   the CTA-N already being built with PMT cameras. However,   for
CTA-S, as well as for a possible upgrade of the CTA-N cameras,   the
LST consortium is currently investigating an upgrade to
SiPM~\citep{Rando:2015jpa,Arcaro:2017owo}. In such a case, the SiPM
photon detection efficiency (PDE)   at the LGS wavelength would
be about four times that of the LST PMT. The LST with SiPM-equipped camera does, however, not have a
technical design implementation plan, and is not yet approved. A simpler
solution could be that of replacing each PMT with an array of SiPM
matrices joined together, in this case the pixel FOV will not
change. Different solutions with different pixel sizes and light-guides
are under discussion and will not be treated here further.

\noindent
Following their use in \autoref{eq.final}, we 
combine the effective telescope dish size, the pixels' field-of-view and their PDE into a new parameter labelled ``LGS sensitivity'', shown 
 in the last column of \autoref{tab2}. One can see that the acceptance of laser
track light in an LST camera pixel results to be a bit more than (or a factor ten higher in the case of a SiPM-based LST camera) that of
an MST camera pixel. Some of the MSTs, \re{however}, approach the LGS
much more than the LSTs do. 
%
Because both MST and LST will have very similar (or
even the same) super-bialkali photo-multipliers, at least before a
possible upgrade, the differences in $\pde$ should be negligible
between both. Values of $\pde = (0.06\pm 0.01)$ can be expected,
maximally varying from 5\% to 9\% \citep[see, e.g.,][]{Mirzoyan:2017,TOYAMA2015280}. 
\re{On the other hand}, the SST
cameras~\citep{Maccarone:2017,Samarai:2017}, equipped with SiPM, are very sensitive at 589~nm, of the
order of 30\%~\citep{Billotta:2014,Otte:2017}.
However, some of the current SST designs try to cover the camera
with protective windows coated with an optical filter to remove
wavelengths longer than 550~nm. 

\begin{figure*}
  \centering
  \hspace{0.05\linewidth}
  \includegraphics[width=0.32\linewidth]{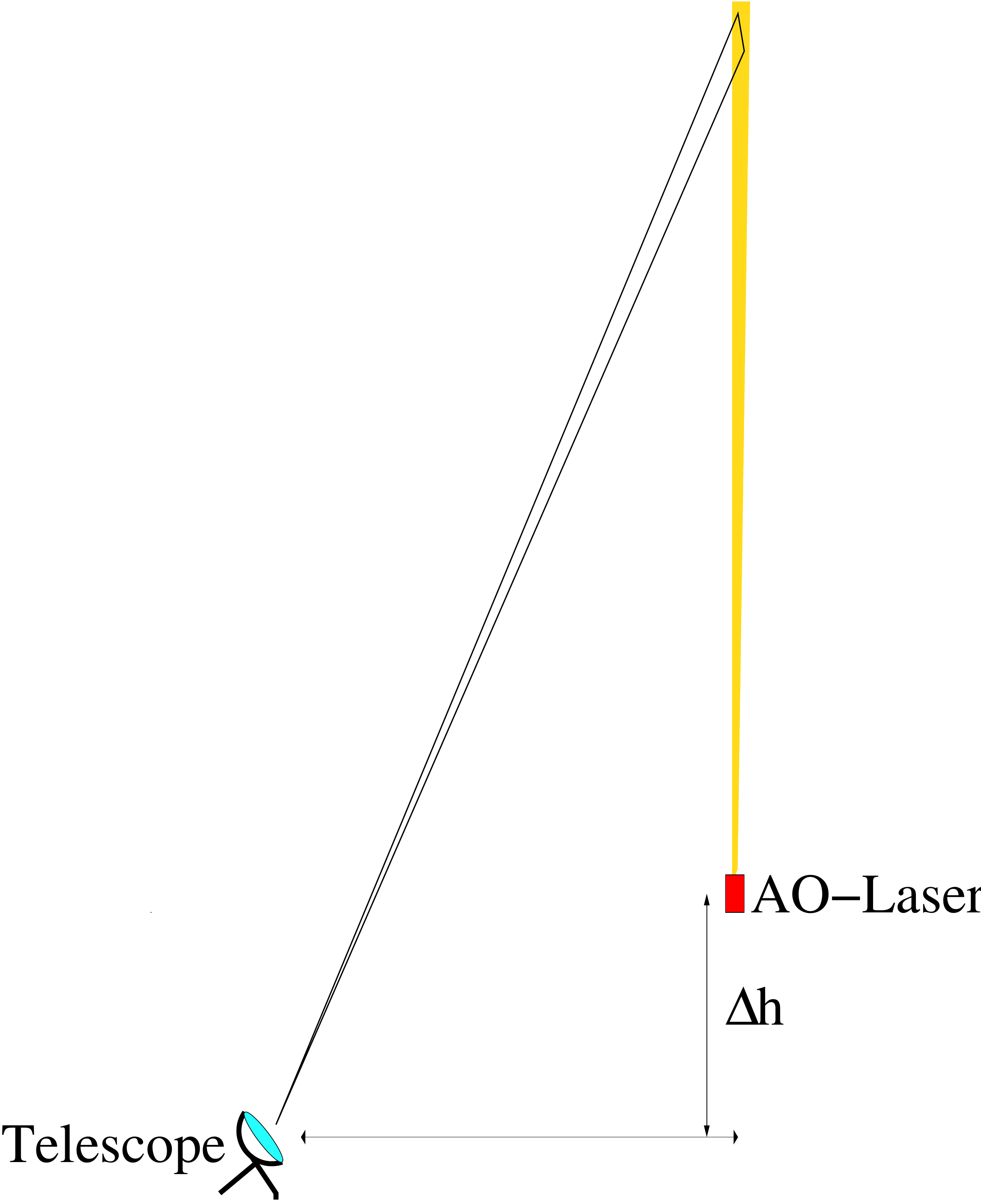} \hspace{0.15\linewidth}
  \includegraphics[width=0.42\linewidth]{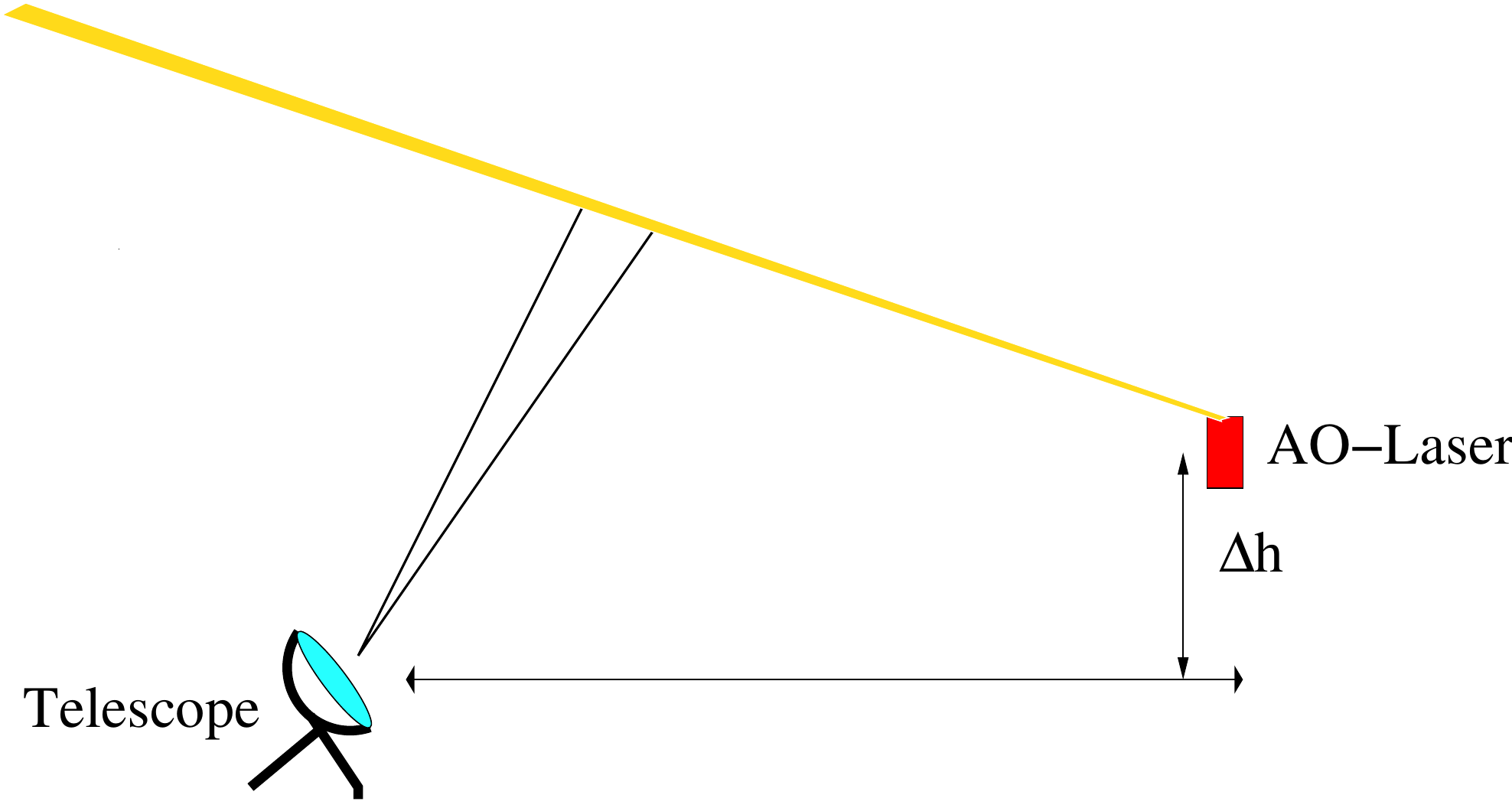} 
  \includegraphics[width=0.48\linewidth]{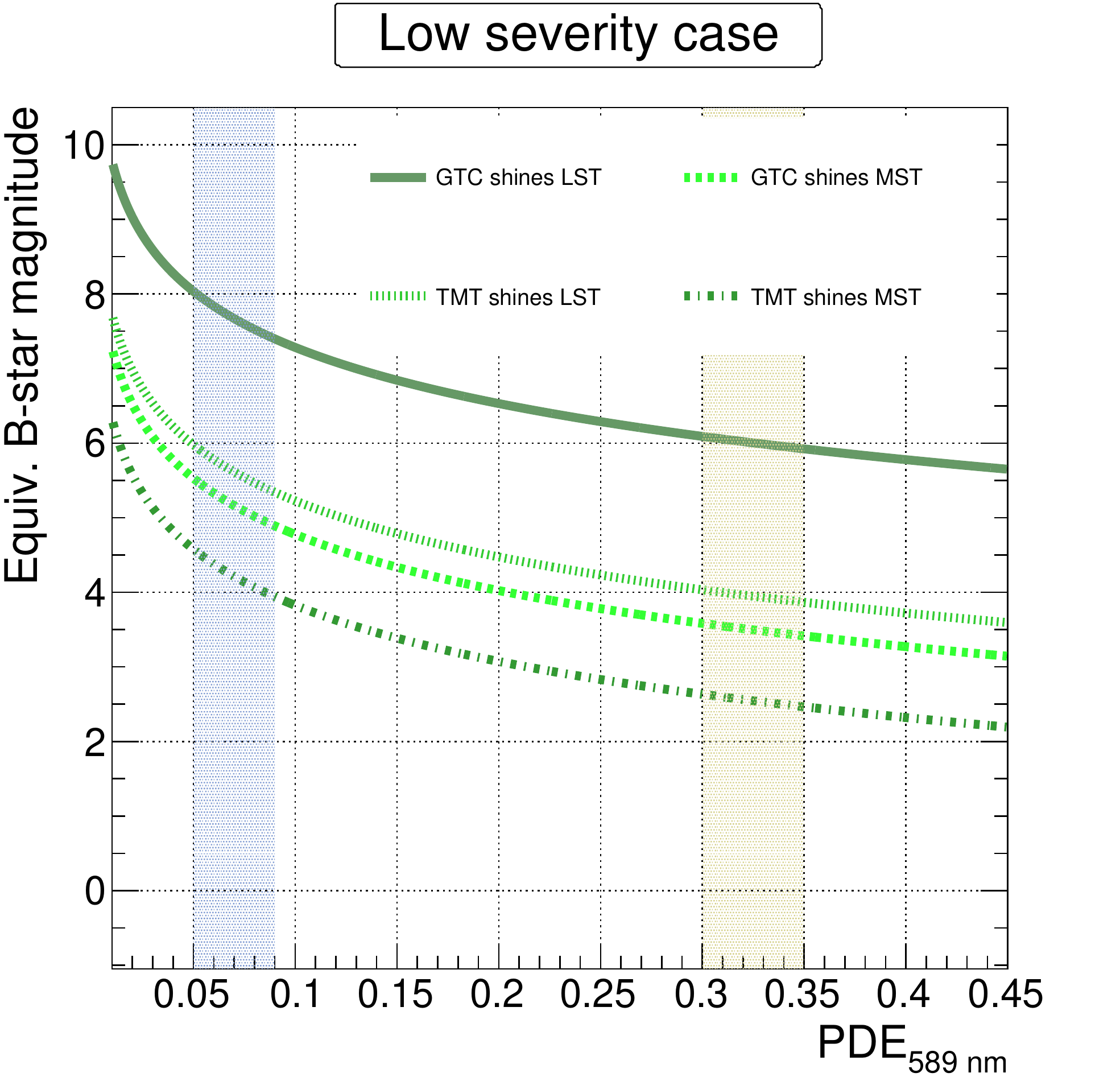}
\includegraphics[width=0.48\linewidth]{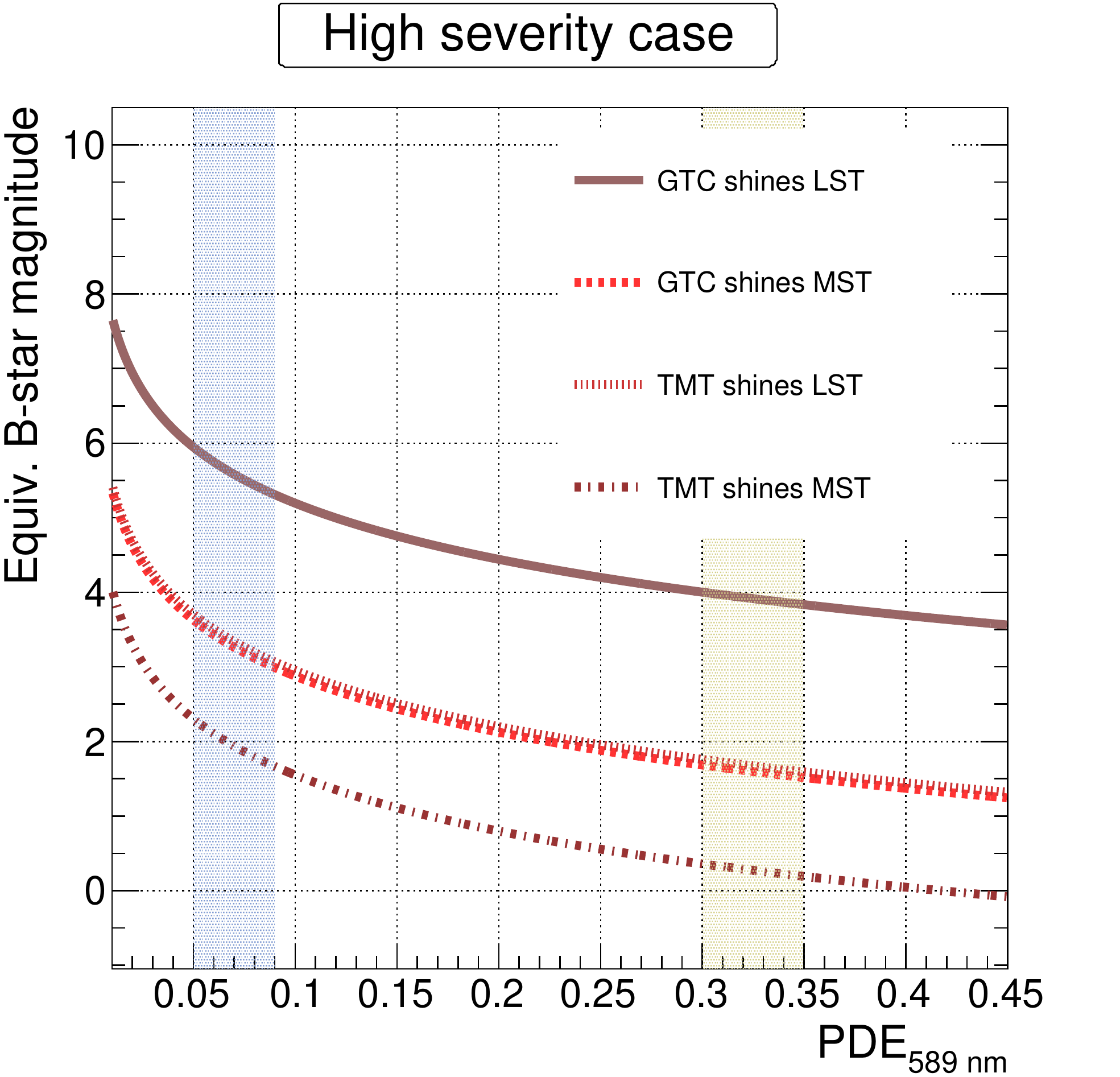}
\caption{\label{fig2} \re{Left: a low severity case scenario, with the LGS laser pointing upwards, while
  the CTA-N observes at 45$^\circ$ zenith angle towards the LGS.
  Right:  a maximal severity case scenario, with the
  laser pointing towards CTA-N, under the lowest allowed elevation,
  and CTA-N looking into the laser beam.
  Top: Sketch of the pointing situation,   }
  \protect Bottom: Equivalent
  $B$-star magnitudes for camera pixels for different PDE values at
  589~nm in these case scenarios. \protect The blue and yellow shaded
  areas depict the case for PMT, and SiPM-equipped cameras,
  respectively. The magnitudes have been derived assuming the
Vega spectrum from~\protect\citet{Bohlin:2007} and a PDE at 440~nm wavelength
of $\mathrm{PDE}_{440\,\mathrm{nm}}=0.35$ and a spectral width of the
PDE of $\mathrm{d}\lambda/\lambda=0.2$.  A global atmospheric
extinction of $z=0.25$ has been further assumed for the star light at
$B$-filter wavelengths. } 
\end{figure*}

In order to provide reference numbers illustrating the severity of a
laser beam crossing the CTA telescopes' field-of-view, we have
selected two cases scenarios.
Concentrating on the relative direction of laser and telescopes, we
define: 
\begin{enumerate}
\item A {\bf low severity case}, expected to happen most frequently
  among the presented scenarios: The CTA observes at 30$^\circ$ zenith
  angle towards the South, i.e., the LGS systems at the ORM, or the North, i.e. the LGS at the \re{ELT}, whose lasers point vertically
  upwards, and the beams cross (see \autoref{fig2} left). The
  distance to the laser beam is then always larger than 500~m at the
  CTA-N, and larger than 30~km at the CTA-S.  
  Scattering  occurs then at altitudes higher than 500~m in the North,
  and 25~km in the South, respectively. In this case, the scattering
  angle is 150 degrees and scattering normally dominated by molecules. 
\item A {\bf maximal severity case}, yielding the highest possible impact
  of the lasers on a CTA telescope, although this scenario is very
  unlikely to occur: The laser \re{propagates} at the minimally allowed
  elevation exactly towards the CTA, where the telescopes look into
  the direction of the laser, at 65$^\circ$ elevation (see
  \autoref{fig2} right).  
  %
  The distance to the laser beam is then as low as 150~m in the
  case of the closest MST to the GTC laser beam, observed at only
  130~m altitude above the MST. 
  Scattering occurs at altitudes ranging from 130~m to 480~m for the closest MST and LST, respectively. 
  In this case, the scattering angle is 90$^\circ$
  and scattering of the laser light is likely dominated by aerosols.  
\end{enumerate}

The results of these case scenarios are quantified in
Table~\ref{tab3}. All of the studied cases focus the beam size into one camera pixel $\Omega_\mathrm{blur} = 1$, even in the extreme case of a maximum approach of the
TMT lasers. In that case, the laser will have a width of about $w \approx 0.3$~m, observed at a distance of $D \approx 400$~m, hence $w/D <  0.8$~mrad$\ll \fovp$. 

As a first important outcome, we see that both in the low and maximal severity case, 
the impact of either the VLT or \re{ELT} LGS is negligible in the Southern Hemisphere installation: 
the predicted photo-electron rate is always below 0.3~p.e./ns for the cameras equipped with PMTs. 
This value is of the order of the p.e. rate produced by the local Night Sky Background light, expected to
produce roughly $0.3\div0.4$~p.e./ns~\citep{Fruck:JI2015a}. 
Such a small effect, limited to few pixels, is properly treated in the data reconstruction. 
\re{SiPM equipped SST cameras expect a Night Sky Background rate of $\gtrsim 0.04$~p.e./ns. The LGS induced may produce an additional background rate of the same order
  of magnitude, however these cameras are not at all limited by such negligible backgrounds.}

A different result is instead found for the Northern Hemisphere installation of the CTA. 
The lasers from both the TMT and the GTC do have a sizeable impact on the camera images of the CTA telescopes, 
even in the low-severity case.  
In order to further illustrate the results for the CTA-N, we compare
the LGS-induced photo-electron rates with those expected from a star
illuminating the same pixel. \autoref{fig2} (bottom) shows the equivalent
$B$-star magnitudes in one pixel vs. the photon sensors PDE at the
LGS wavelength for the CTA-N. Two vertical bands highlight the
PDE of PMT-like sensors (left blueish band) and SiPM-like sensors
(right yellowish band). One can observe that the laser light may
produce the same photo-electron rate as that of a $B$-star of magnitude
$1^m\div2^m$ in the maximal-severity case. 

%
%
%
%
%
%
\begin{table}
\centering
\begin{tabular}{l|cccccc|c}
  \toprule
  \rowcolor{Gray}
  Telescope & Area   & Pixel FOV    & Camera FOV & \multicolumn{3}{c}{PDE}     &  {\bf LGS Sensitivity} \\
  \midrule
  \rowcolor{Gray}
           & $\atel$ & $\fovp$      &            & $\pde$ & \re{($\textit{PDE}_{515\,\mathrm{nm}}$)} & \re{($\textit{PDE}_{355\,\mathrm{nm}}$)} & $\fovp \cdot \atel \cdot \pde $  \\ 
  \rowcolor{Gray}
           & (m$^2$) & (mrad)       &   (deg)    &         &    &    & $(\mathrm{m}^2 \cdot\mathrm{rad})$    \\
  \midrule
  LST-PMT  &  370    & 1.75         & 4.3        &  $0.06 \pm 0.01$         & $0.20 \pm 0.03$         &  $0.42 \pm 0.03$         &  $\sim$0.039   \\       
  LST-SiPM &  370    & 1.75         & 4.3        &  $0.28 \pm \dagger$      & $0.37 \pm \dagger$      &  $0.3 \pm \dagger$      &  $\sim$0.155   \\
  MST      &   88    & 3.0          & 7.6        &  $0.06 \pm 0.01$         & $0.20 \pm 0.03$         &  $0.42 \pm 0.03$         &  $\sim$0.016   \\
  SC-MST   &   41    & 1.2          & 7.6        &  $0.28 \pm \dagger$      & $0.37 \pm \dagger$      &  $0.3 \pm \dagger$      &  $\lesssim$0.012 \\ 
  SST-DC   &  7.5    & 4.2          & 8.8        &  $0.28 \pm 0.10^\ddagger$ & $0.32 \pm 0.10^\ddagger$  & $0.3 \pm 0.05^\ddagger$ & $\lesssim$0.010  \\      
  SST-SC   &  (8.0--8.3)$^*$ & (3.0--3.5)$^*$ & (8.3--10.5)$^*$ &  $0.28 \pm 0.10^\ddagger$ &  $0.32 \pm 0.10^\ddagger$ &  $0.3 \pm 0.05^\ddagger$ & $\lesssim$0.009  \\      
  \bottomrule
\end{tabular}
\caption{Characteristics of the different CTA telescope types. \re{For completeness, the PDE is not only displayed at the canonical wavelength of 589~nm, but also the pulsed Rayleigh laser systems
    operating at 355~nm and 515~nm \protect\citep{Tokovinin:2016,Rutten:2006}.}\label{tab2}\newline
$\dagger $ The SiPM development for LST (and SC-MST) pixels is still ongoing,
  uncertainties can be as large as 50\%.
$\ddagger$These cameras will probably be covered by a window coated
    with an optical filter which cuts out wavelengths
    $>$550~nm. 
$^*$Actual values depend on concrete implementation. } 
\end{table}

\begin{table*}
  \centering
  \begin{tabular}{r@{$~\rightarrow~$}l|D{.}{.}{3.0}D{.}{.}{3.0}rl|D{.}{.}{3.0}D{.}{.}{3.0}}
    \toprule
      \rowcolor{Gray}
    \multicolumn{2}{l|}{Case}       & \multicolumn{1}{c}{Distance to} &  
    \multicolumn{1}{c}{Scattering} &   \multicolumn{1}{c}{Flux per
      pixel}  & Main scattering factor & \multicolumn{2}{|c}{{\bf Flux per pixel (N=6)}}     \\
      \rowcolor{Gray}
    \multicolumn{2}{l|}{}           & \multicolumn{1}{c}{laser track} &
    \multicolumn{1}{c}{Altitude  } &  \multicolumn{1}{c}{(for $N$
      lasers)}  &  &  \multicolumn{1}{|c}{{\bf PMTs}}  &
    \multicolumn{1}{c}{{\bf SiPMs}}      \\
      \rowcolor{Gray}
    \multicolumn{2}{l|}{}           &   \multicolumn{1}{c}{ (km)   } &
    \multicolumn{1}{c}{ (km)     } & \multicolumn{1}{c}{ (p.e./ns)}  &
    &     \multicolumn{1}{|c}{{\bf(p.e./ns)}}  & \multicolumn{1}{c}{ {\bf(p.e./ns)}}      \\
    \midrule
    \rowcolor{Gray}
    \multicolumn{8}{l}{{\bf CTA Northern Hemisphere Site}} \\
    \midrule 
    \multicolumn{8}{l}{Maximal Severity} \\
    \midrule 
    GTC & LST & 0.30 & 0.27 & \re{76} $\cdot \pde $ & Aerosol & 5  & \re{24} \\
    GTC & MST & 0.15 & 0.13 & \re{76} $\cdot \pde $& Aerosol & 5  &  \multicolumn{1}{c}{n.a.}    \\ 
    TMT & LST  & 0.53 & 0.48 & \re{34} $\cdot N \cdot \pde $ & Molecular/Aerosol & \re{12} & \re{65} \\
    TMT & MST  & 0.39 &  0.35 & \re{22} $\cdot N \cdot \pde $& Molecular/Aerosol & \re{8}  & \multicolumn{1}{c}{n.a.} \\
    \midrule
    \multicolumn{8}{l}{Low Severity} \\
    \midrule 
    GTC & LST & 1.10 & 0.95 & \re{29} $\cdot \pde $& Molecular &  2  & \re{9} \\
    GTC & MST & 0.52 & 0.45 & \re{30} $\cdot \pde $& Molecular/Aerosol &  2  &  \multicolumn{1}{c}{n.a.}    \\
    TMT & LST  & 2.30 & 2.00 & \re{12} $\cdot N \cdot \pde $& Molecular  & \re{4} &  \re{22} \\
    TMT & MST  & 1.64 & 1.42 & \re{7} $\cdot N \cdot \pde $& Molecular  &  3  &  \multicolumn{1}{c}{n.a.}\\
    \midrule
    \rowcolor{Gray}
    \multicolumn{8}{l}{{\bf CTA Southern Hemisphere Site}} \\
    \midrule
    \multicolumn{8}{l}{Maximal Severity} \\
    \midrule 
    \re{ELT} & LST  & 7.4 & 6.8 & \re{0.6} $\cdot N \cdot \pde $& Molecular  &  0.2  & 1.2 \\
    \re{ELT} & MST  & 7.2 & 6.5 & 0.3 $\cdot N \cdot \pde $& Molecular  & 0.1 &  \multicolumn{1}{c}{\quad ~~ n.a.}\\
    \re{ELT} & SST  & 7.0 & 6.3 & \re{0.03} $\cdot N \cdot \pde $&  Molecular  & \multicolumn{1}{c}{\quad ~~ n.a.}  & 0.06  \\
    \midrule 
    \re{VLT}& LST& 4.8 & 4.4 & 1.2 $\cdot 4 \cdot \pde $& Molecular  &  0.3  & 1.6 \\
    \re{VLT}& MST& 4.5 & 4.1 & 0.5 $\cdot 4 \cdot \pde $& Molecular  & 0.1 &  \multicolumn{1}{c}{\quad ~~ n.a.}\\
    \re{VLT}& SST& 4.3 & 3.9 & 0.07 $\cdot 4 \cdot \pde $&  Molecular  & \multicolumn{1}{c}{\quad ~~ n.a.}  & 0.09  \\
    \midrule 
    \multicolumn{8}{l}{Low Severity} \\
    \midrule 
    \re{ELT} & LST  & 31.4 & 27.2 & \re{0.06} $\cdot N \cdot \pde $&  Molecular  & 0.02  & 0.1 \\
    \re{ELT} & MST  & 30.2 & 26.2 & 0.03 $\cdot N \cdot \pde $& Molecular  & 0.01  & \multicolumn{1}{c}{\quad ~~ n.a.} \\
    \re{ELT} & SST  & 29.2 & 25.3 & 0.004 $\cdot N \cdot \pde $&  Molecular  & \multicolumn{1}{c}{\quad ~~ n.a.}  & 0.007  \\
    \midrule 
    \re{VLT}& LST& 20.6 & 17.8 & 0.2 $\cdot 4 \cdot \pde $&  Molecular  & 0.06  & 0.3 \\                            
    \re{VLT}& MST& 19.4 & 16.8 & 0.1 $\cdot 4 \cdot \pde $& Molecular  & 0.03  & \multicolumn{1}{c}{\quad ~~ n.a.} \\    
    \re{VLT}& SST& 18.6 & 16.1 & 0.02 $\cdot 4 \cdot \pde $&  Molecular  & \multicolumn{1}{c}{\quad ~~ n.a.}  & 0.02  \\
    \bottomrule
  \end{tabular}
  \caption{Expected photo-electron fluxes in a camera pixel for the two studied severity cases. \label{tab3}}
\end{table*}

\re{If both LGS laser and CTA observe the same source, the CTA cameras will observe the \ree{Rayleigh plume from an altitude $> 2D/\fovc$ and} fluorescence emission of the mesospheric sodium layer.
  In the worst case, the illuminated sodium layer will be seen by CTA under an angular length $\Psi$ of: }
\begin{equation}
   \Psi \lesssim \frac{\Delta H \cdot L}{H^2}  \quad,
  \end{equation}
\noindent
\re{where $H$ denotes the distance to the layer centroid $H\approx 89.7$~km and $\Delta H$ the average layer width.
  The resulting angular length is always smaller than one camera pixel for the CTA-N. The full layer will hence be seen as just an additional star.
In the South, $\Psi$ can become as long as 10 camera pixels, however their average flux results to be always less than 0.05~p.e./ns, even in the case of an upgraded LST camera. }   
\ree{The photo-electron rate from the Rayleigh plume, observed by the outmost camera pixel at the CTA-N
  is visible in the closest MST from greater than 4$\div$12~km for the GTC and TMT lasers, respectively, and the received rate is always smaller than}
\ree{1.6~p.e./ns for both cases at 60$^\circ$ observation zenith angle and $<$0.9~p.e./ns for observations at zenith, using Eq.~\ref{eq:parallel}.
  In the absolutely worst case, the Rayleigh plume will leave spurious photo-electron rates larger than those from the typical night sky background in a line starting from the outer camera edge up to half the camera radius. }.

\subsection{Probability of interference during CTA fast repositioning}
\label{sec:results_prob}

In this section we make use of the formalism of \autoref{sec.prob} to
estimate the fraction of time in which the LGS will interfere with CTA
operations in such a way that the underlying science case may be degraded or put at risk.
Given the results from the previous section, in which we show that
both the VLT and \re{ELT} LGS will have a negligible impact on CTA-S,
we will focus on the CTA-N only, and particularly on the interference of the TMT LGS with the CTA telescopes. 
We cannot make accurate predictions of the foreseen observing programs of the CTA-N, for the time after the TMT will start operations\footnote{foreseeably after 2027.},
because most of the CTA observing time will be open for guest observer proposals. However, the CTA consortium can use 40\% of the first 10 years of CTA operations in the form of proprietary key science projects. 
We use these to make a reasonable guess as to the distribution of target and observation types. 

Some of the CTA-N's core science deals with fast transients
and amount to about
45~hr/yr/site for galactic ToO's and
120~hr/yr/site for
extra-galactic ones for the first 12 years of operation~\citep[see chapter 9 of][]{2017arXiv170907997C}.
We did not include open time, nor director's time, which may increase
that number further. Both are however not expected to alter the
previous numbers significantly. A prediction for the time reserved
for ``rapid'' multi-wavelength campaigns with allocated and
immovable time slots can be obtained from chapter~12 of \citet{2017arXiv170907997C} summing up to 
245~h/yr for the CTA-N, all dedicated to extra-galactic targets.  
%
Further assuming that CTA-N follows up each alert for an average of
two~hours (which is rather standard for this technique),  
we can expect about
180
fast or immovable re-positionings per year,  
hence one fast or immovable re-positioning every one and a half nights for extra-galactic
targets and one fast re-positionning every
11 nights for galactic sources. 

 \begin{table}
    \centering
    \begin{tabular}{lcc}
      \toprule
      Opt. Telescope & Closest CTA-N  &    $\delta_\mathrm{tel}$ \\
      &   telescope    &     (deg.)     \\
\midrule
GTC   &    LST       &    0   \\
      &    MST       &   -22  \\
TMT   &    LST       &    22  \\
      &    MST       &    20  \\
\bottomrule
\end{tabular}
\caption{\label{tab:angles}Tilts of the lines connecting an optical telescope location with a CTA-N telescope, with respect to the North-South axis. }
\end{table}

Fast ToO observations are typically expected to happen in \textit{normal observing mode}, i.e. making use of the typical CTA-N field-of-view of about eight degrees\footnote{which is, in this case, provided by the MST cameras, while for CTA-S a larger FOV of ten degrees is obtained with the SSTs,
  see \autoref{tab2}. The CTA provides also the possibility to observe with even larger fields-of-view of up to 15$^\circ$ in ``divergent pointing mode''. These are, however, not expected to be employed for
 rapid pointings, at least for the moment. }. 
However, we may assume that such observations may be observed up to one degree off-axis\footnote{%
The sensitivity of IACT telescopes decreases off the optical axis of the telescopes, however this happens rather slowly, as shown for the MAGIC
case~\citep{Aleksic:2014lkm}, and in CTA in recent simulations. The latter show that  
maximally 20\% loss of point-source sensitivity are obtained for the case of one degree off-pointing, 
considerably reduced for medium and high gamma-ray energy ranges.},
to avoid the laser beam, hence $\textit{FOV}_\mathrm{vetoed} \approx  6^\circ$.

To start, we set $\rcrit$ to the typical night sky background \re{(NSB)} rate for extra-galactic sources. 
This somewhat arbitrary criterion has been chosen assuming that the individual pixel rate control will get active, at least in the case of the LSTs, and raise the trigger thresholds of the illuminated pixels. 
Loss of sensitivity at the energy threshold is then expected. Later on, we will investigate in more detail the dependency of the observation time loss on $\rcrit$.
The maximum altitude $\Hm$ comes out to be approximately $14\div 20$~km above ground for the GTC laser and $18 \div 26$~km above ground for the six TMT lasers used together. 
If the LST camera is equipped with SiPMs, without further protecting filters, the LGS
light will disturb observations up to 32~km above ground.

We assume now a maximum zenith angle for CTA-N's rapid re-positionning targets of $\thetam = 45^\circ$. 
The extra-galactic ToO occurrence phase space covers then a solid angle of about $\Omega_\mathrm{ToO,extra-gal}\approx 1.8$~sr, and $\Omega_\mathrm{ToO,gal} \lesssim 0.3$~sr. 

\begin{figure*}
\centering
\includegraphics[width=0.99\linewidth]{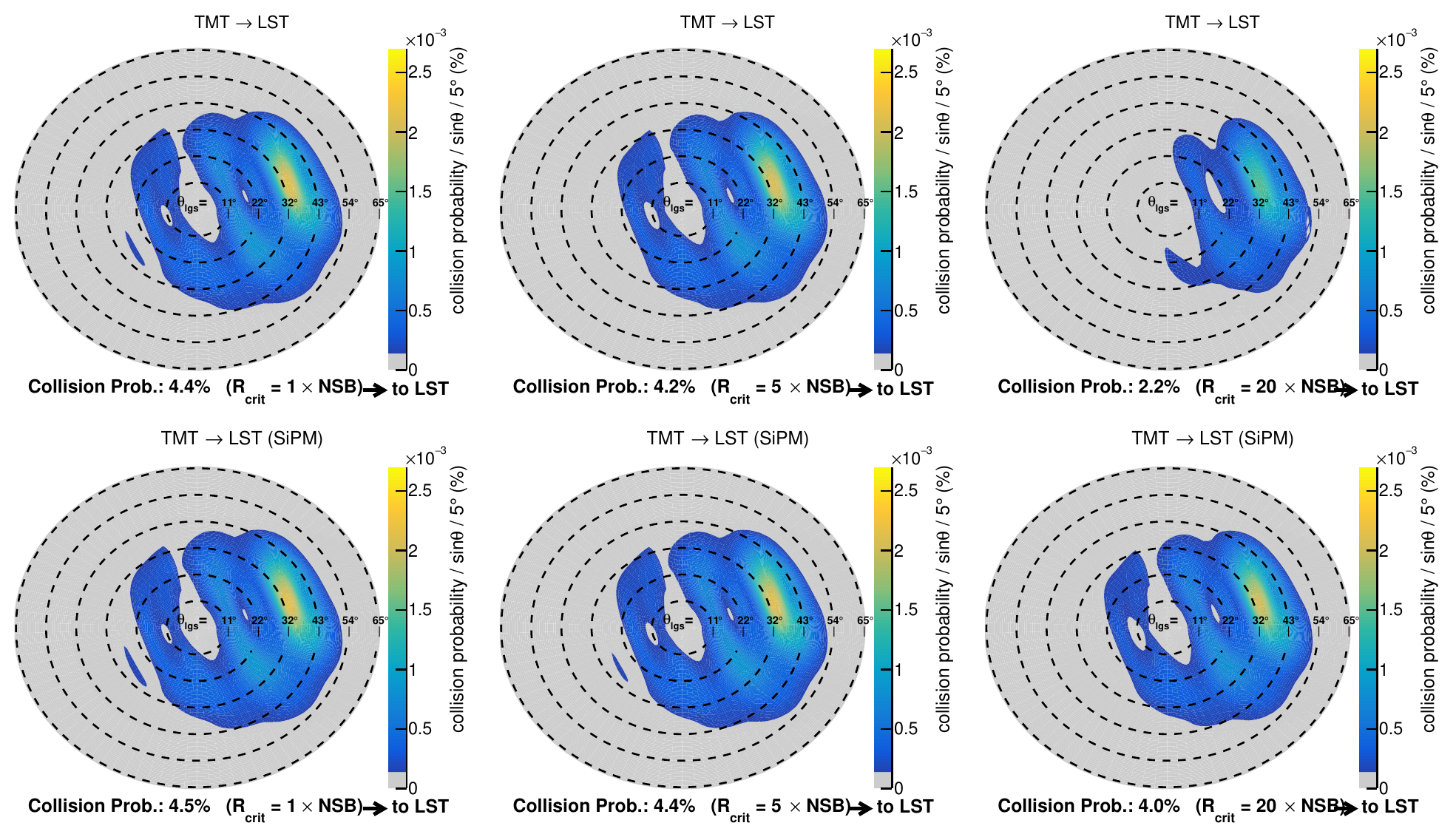}
\caption{Probability map of conflicts of the LGS with the observability region for extra-galactic CTA-N ToO's, depending on the LGS-lasers' local pointing angles $(\thetal,\phil)$, obtained 
from a smoothed  map obtained from 10 years of MAGIC data pointing history in local zenith/azimuth coordinates
 (upper two figures: from 0 to 60 deg, center and bottom figures: from 0 to 65 deg). 
 The laser is located in the center, the direction to the CTA-N telescope points always towards its right side, 
but the different pointing probabilities for the LGS (with respect to CTA-N) have been taken into account. 
The text on the bottom shows the total integrated probability. If instead the pointing probability map from one year of GTC pointing is used, 
conflict probabilities about half a percent higher are obtained.
\label{fig.probcollision}} 
\end{figure*}

We apply the above occurrence estimates of fast ToO's and time slots reserved for multi-wavelength campaigns, together with the distances between CTA-N telescopes and the two LGS facilities at GTC and the TMT (\autoref{tab:laser_params})
and the values of $\delta_\mathrm{tel}$ (\autoref{tab:angles})
to \autoref{eq.pconflict}. 
The resulting probability maps to reside in a certain pointing direction times the 
probability to veto a CTA-N ToO pointing in that direction for extra-galactic observations (the so-called "conflict probability maps") 
are shown in \autoref{fig.probcollision} (left side), for 
the LST case only, but are almost identical for the case of the MST. The reason for this similarity can be found in the geometry of the system: 
because the maximum altitudes $\Hm$ are considerably larger than the distance between LGS and CTA telescope in both cases 
and consequently the intersection heights $h$, the contribution of the visible laser path length at the highest altitudes (e.g. from 20 to 26~km) 
to the vetoed angular length $\Psi$ is small. In other words: $\Psi$ scales in zero'th order as $\atan(\Hm/L)$, which becomes flatter and flatter as 
the argument $\Hm/L$ gets larger. The vetoed pointing maps for CTA-N are hence rather insensitive to the exact values of $\Hm$ and consequently $\rcrit$.
\re{Even assuming $\rcrit=0$ (i.e. no collision allowed at whatever level), the conflict probabilities increase by less than 5\% with respect to $\rcrit=1$.}
In order to highlight these dependencies, we show the same conflict probability maps for largely enhanced values of $\rcrit$ for the case of 
an LST camera equipped with PMTs (top) or with SiPMs (bottom). As expected, the SiPM equipped telescope show higher conflict 
probabilities with a smaller dependency on $\rcrit$, whereas the PMT equipped camera can reduce the conflict probability by about a factor of two, 
if $\rcrit$ is chosen to be twenty times larger than the typical background rate.
The integral of all maps are written below on the same figures and provide the total probability to have an 
extra-galactic CTA-N ToO vetoed, \textit{once the LGS is used at all}.  
For comparison, we also checked the local pointing field of one year of GTC pointing (courtesy of Antonio Luis Cabrera Lavers) and found compatible results.
\re{If we use instead a one-year VLT LGS pointing history together with the MUSE instrument, about 20\% lower conflict probabilities are obtained.}

The integrated probabilities are finally inserted in \autoref{eq.pcollision}, using the LGS duty cycles listed in \autoref{tab:laser_params}.
These final results are summarized in \autoref{tab:probtot}.

\begin{table*}
\centering
\begin{tabular}{lcccccc}
\toprule
Optical     &  Duty cycle  & Max. laser &  Prob. conflict & Prob.     & Estimated  & Estimated  \\
 Telescope  &  LGS         & zenith     &  per simult.    & conflict  & number of  & occurrence \\
            &  $\eta$      & angle      &  observations   & per ToO   & ToO's      & of conflict  \\
            &   (1)        & (deg.)     &     (1)         &   (1)     & (yr$^{-1}$) & (yr$^{-1}$) \\
\midrule
\addlinespace[0.15cm]
\multicolumn{6}{c}{Extra-galactic ToO's CTA-N}  \\[0.3cm]
GTC          &  0.15         & 60 &  0.045 & 0.007 & 180 & 1.2 \\
TMT          &  0.75         & 65 &  0.045 & 0.034 & 180 & 6.1 \\
\addlinespace[0.15cm]
\midrule
\addlinespace[0.15cm]
\multicolumn{6}{c}{Galactic ToO's CTA-N} \\[0.3cm]
GTC galactic      &  0.08  & 60 &   0.094 & 0.007 &  22 & 0.17  \\
GTC extragal.     &  0.07  &    &   0.045 & 0.003 &  22 & 0.07  \\
GTC total         &  0.15  &    &         & 0.008 &     & 0.24  \\
\addlinespace[0.15cm]
TMT  galactic      &   0.38  &65 &   0.094  & 0.036 & 22 & 0.78  \\
TMT  extragal.     &   0.37  &   &   0.045 & 0.017 &  22 & 0.37  \\
TMT  total         &   0.75   &   &         & 0.053 &    & 1.05  \\
\bottomrule
\end{tabular}
\caption{\label{tab:probtot}Probabilities of conflicts (``collisions'') between LGS and CTA-N fast-ToO observations. }
\end{table*}

\section{Summary and Conclusions}
\label{sec:discussion}

In this work, we have explored the effect that Laser Guide Star (LGS) facilities,
  as those foreseen on several present and future large optical telescopes, such
  as the \re{VLT, ELT}, the GTC and the TMT, have on neighbouring telescopes, particularly those observing with large FOVs.
  LGS systems operate high power continuous wave lasers  
  at \re{589.159}~nm \ree{vacuum} wavelength to excite sodium nuclei in the upper mesosphere. 
  The laser light can scatter into the FOV of the neighbouring instruments  
  and affect data taking or reconstruction. 

  We have computed general equations to predict
  the number of scattered photons into a camera pixel, 
  as well as estimates for the fraction of
  time lost because of possible 
  crossings of the neighbouring telescope FOV by the 
  laser beam. We have later on applied those
  equations quantitatively to the case of the CTA, a planned ground-based array of 
  gamma-ray instruments, currently under construction at the
  Observatorio del Roque de Los Muchachos, La Palma, and soon at the Armazones valley, close to Paranal, in Northern
  Chile. The Northern Hemisphere array, CTA-N, will contain two types of telescopes, the LST and the MST, 
  and may be affected by the GTC, and possibly the TMT, LGS. 
  In the Southern Hemisphere, the CTA-S will contain three types of telescopes, adding the SST type, 
  and is located close to the \re{VLT and ELT} LGS.  

  In 
  \autoref{sec:scenarios}, the amount of scattered 
  laser light into  CTA camera pixels has been computed, and two case
scenarios, a low and a maximal severity case, studied. 
The obtained numbers provide a rough estimate for the ranges within which 
LGS induced photo-electron rates can be expected, namely
for the GTC laser: from 2--5~p.e./ns for PMT-based LST or MST cameras to
  about 10--30~p.e./ns. for an LST camera equipped with SiPM; for the
  TMT LGS with six simultaneous beams: from 3--14~p.e./ns. for PMT-based LST or MST cameras to
  about 30--80~p.e./ns. for an LST camera equipped with SiPM; for the more distant
  \re{VLT with four lasers or} \re{ELT} lasers with six beams, rates lie always well below 1.6~p.e./ns, even in the case of an LST camera equipped with SiPM, otherwise below 0.3~p.e./ns.
  The SSTs, which are only deployed in the South, are not affected at
  all by the \re{VLT or the} \re{ELT} lasers (rates below 0.01~p.e./ns),
  although they approach the \re{VLT and} \re{ELT} most.
The critical combination is hence the one of the TMT (in less extent
the GTC) lasers shining into the CTA-N telescopes, especially if LST cameras are potentially upgraded to SiPM in the future.
The obtained count rates can be compared to those of the night sky background, expected to
produce roughly 0.3(0.4)~p.e. per MST (LST) pixel per nanosecond,
respectively~\citep{Fruck:JI2015a}. However, observations are also
planned under partial moon light, with night sky background rates up
to about 20 times higher than the previous numbers, under reduced sensitivity~\citep[see e.g.][]{magicmoon}.
The effect of all investigated cases is similar to having a row of magnitude down to as low as $1^m$,
B-stars crossing the camera.
Even if the six TMT lasers are fired in divergent mode, only one row
will be seen in the telescope cameras, i.e. the different laser beams
cannot be resolved.

\re{The fluorescing sodium layer itself measures 11~km on average, and can even reach 16~km in exceptional cases~\citep{Moussaoui:2010}.
    It is however harmless if found in the FOV of the neighbouring telescope, if both installations are sufficiently close (as is the case at CTA-N).
    At larger distances, the layer can spread over several pixels (as at CTA-S), but the light flux received by a single camera pixel is then considerably lower than the artifical star produced in the telescope housing the LGS,
    and probably negligible. This is the case at CTA-S. }.
\ree{Some residual spurious light of the order of 1~p.e./ns will be received, however, at the CTA-N by the outmost camera pixels of the closest MSTs from the Rayleigh plume of the LGS lasers of the GTC and the TMT, if both installations observe the same source.}
  
In this situation, the laser photons  are not a
danger for the safety of the CTA-N cameras. Each CTA-N camera pixel is
equipped with an automatic  high-voltage down-regulation in case of
excessive anode current, which makes observations safe. 
This mechanism will probably also
protect the CTA-N against too high data rates of fake triggers. 
However, analysis of data affected by such an LGS laser beam crossing the camera is challenging and should be avoided, apart from the inevitable loss of sensitivity.
Experience with the MAGIC telescopes has shown,  that additionally the laser beam can confuse the star-guider analysis software, used to correct the 
pointing of the telescopes with the help of CCD cameras.  
Whereas solutions based on Notch-filters~\citep{Schallenberg:2010} exist for the CCD cameras, similar approaches for the CTA camera pixels require
future study and some innovation effort: coating of curved surfaces with filters is not straight-forward, nor thin filters resisting all types of weather
phenomena, like temperature changes, humidity cycles, etc. to which, for instance, a protecting plexiglas of the cameras is exposed during night. 
Also losses of Cherenkov photons in the wavelength range from 300~nm to 550~nm are an issue.

In 
\autoref{sec:results_prob}, we derived the probabilities that a CTA-N observation collides with the LGS laser beam causing an 
unacceptably high photo-electron rate in the CTA-N camera, \ree{when different sources are observed\footnote{%
The corresponding code to produce the figures, written in \textit{ROOT}, is available on demand.
    }.}
We find around 1\%(3\%) for extra-galactic observations of CTA-N to collide with the GTC(TMT) LGS beams, 
and around 1\%(5\%) for galactic observations, respectively. 
The lower probabilities for the GTC laser are due to both its smaller relative duty cycle and its lower laser power, which in consequence allows to cross 
the CTA-N field-of-view at a lower limiting altitude, even if CTA-N telescopes approach that laser much closer.
These probabilities can be reduced by only 5\%, if LGS laser-induced additional p.e. rates of up to five times the natural dark night sky background
rate are allowed. Relieving this requirement to 20 times the night sky background rate (corresponding to observations under partial moon light), 
reduces the conflict probabilities by about a factor of two, unless a SiPM upgraded LST camera is used, for which the reduction is of the order of 10\% only.

Since both the CTA-N observatory and the GTC/TMT LGS facilities will be included in the Laser Tracking Control System (LTCS) of the ORM, most conflicts can be avoided 
by adequate scheduling of the sources. 
Due to the nature of the current ``basic'' configuration of LTCS at the ORM, which currently follows a  strict ``first-on-target'' policy, 
this is however not the case for fast Target-of-Opportunity (ToO) alerts of CTA-N, which cannot be scheduled to later times 
without putting at risk the science case.


Such fast ToO's, and those requiring simultaneous multi-wavelength or
multi-messenger coverage, will occur 180 (22) times per year for
extra-galactic (galactic) targets, following the key science programs of the CTA-N~\citep{2017arXiv170907997C}.
A collision in such a case will then happen $1\div6$ times a year for
extra-galactic ToO's with the GTC/TMT LGS beam, respectively, 
and $0.2\div1$ times a year for galactic ToO's, \re{excluding those cases where both installations observe the same target, because the glowing sodium layer
  will be imaged  either into one camera pixel (in the case of CTA-N) and hence treated as just an additional star, or become too faint and indistinguable from the night-sky background (in the case of CTA-S).}
In order to minimize the impact of science loss, both for the CTA and for the GTC/TMT,
we suggest a modification of the strict ``first-on-target'' policy of the current configuration of the LTCS.
Such ``enhanced'' versions of the LTCS~\citep{Santos:2016} are already operative at Mauna Kea and Paranal, but require a previous consensus on newly defined priorities for observation targetting. 
Assuming that the relation of time reserved for fast ToO's and multi-wavelength/multi-messenger observations with respect to the total available time is similar to the one
foreseen for the CTA-N, we expect then a reduction of the number of conflicts \re{leading to science loss at one or the other side} by at least a factor of three.
\re{This would include new rules such that the observation of a science target with less urgency by one part yields priority to the other unless the levels of urgency are comparable.}
Such low conflict rates can then be considered negligible, if compared to other, external, disturbances, like technical problems or the weather.

\paragraph*{Acknowledgments}
The authors thank the anonymous referees for their fruitful and professional comments which helped to improve the paper. 
This work has been funded by the grant FPA2015-69210-C6-6-R of the Spanish MINECO/FEDER,~EU. 
The CTA consortium gratefully acknowledge financial support from the agencies and organisations listed at \url{https://www.cta-observatory.org/consortium_acknowledgments}.
This paper has gone through internal review by the CTA Consortium.

\bibliographystyle{mnras}
\bibliography{lgs.bib} 

\begin{appendix}
\label{sec:appendix}

\section{Derivation of distance to the laser beam}
\label{app:distance}

We define the following auxiliary variables:

\begin{align}
   \psi_x  &= \tan\thetal \cdot \cos\phil \quad, \\
   \psi_y  &= \tan\thetal \cdot \sin\phil \quad,   \\
   \zeta_x &= \sin\thetal \cdot \cos\phil \quad,   \\
     A     &= L \cdot \cos\thetal / H \quad, 
\end{align}
\noindent

where $H \cdot \psi_x$ yields $x$ and $H \cdot \psi_y$ yields $y$, and the laser length multiplied with $\xi_x$ yields $x$.

and will later make use of the relations:

\begin{align}
      \psi_x^2 + \psi_y^2 &= \tan^2\thetal \quad,\\
  1 + \psi_x^2 + \psi_y^2 &= 1/\cos^2\thetal  \label{eq.psixpsiy} \quad.
\end{align}  

Applying Pythagoras' theorem on the three triangles contained between the AO-laser, the CTA telescope and the projected beam intersection point on ground, we obtain:
\begin{align}
  L_l^2 &= H^2/\cos^2\thetal \quad,\\
  L_t^2 &= H^2 + l_t^2 \quad, \\
  &= H^2 + (L - H \psi_x)^2 + H^2\psi_y^2 \quad,\nonumber\\
  &= H^2/\cos^2\thetal \cdot \left( 1 + A^2 - 2A\zeta_x \right)   \quad,
  \end{align}
\noindent
where Eq.~\ref{eq.psixpsiy} has been used in the last step,
and $L_t$ and $L_l$ denote the distances from the CTA telescope, or the AO-laser, to the beam intersection point, and $l_t$ and $l_l$ the distances to the projected beam intersection point on ground, respectively.

\section{Derivation of the scattering angle}
\label{app:angle}

We apply the cosine rule to obtain the scattering angle $\vartheta$:
\begin{align}
\cos(\pi-\vartheta) &= \frac{L_t^2+L_l^2-L^2}{2L_tL_l}  \quad, \\
&= \frac{1 - A\zeta_x}{\sqrt{  1 + A^2 - 2 A\zeta_x}} 
\end{align}
\noindent
and obtain:
\begin{align}
  \cos(\pi-\vartheta) &= \frac{1 - A \cdot \zeta_x}{\sqrt{  1 + A^2 - 2A\zeta_x}} \quad,\nonumber\\
  \cos^2\vartheta     &=  \frac{1 + (A\zeta_x)^2 - 2A\zeta_x}{1 + A^2 - 2A\zeta_x} \quad,\nonumber\\
  1 + \cos^2\vartheta &=  \frac{2 + A^2 \cdot (1 + \zeta_x^2) - 4A \zeta_x}{1 + A^2 - 2A\zeta_x} \quad,\nonumber\\
  \sin^2\vartheta &= 1 - \cos^2\vartheta =  \frac{A^2 \cdot (1 - \zeta_x^2)}{1 + A^2 - 2A\zeta_x} \quad,\nonumber\\
  \sin\vartheta\cdot L_t &= A \cdot \sqrt{1 - \zeta_x^2}  \cdot H /\cos\thetal \quad,\nonumber\\
  \sin\vartheta\cdot L_t &= L \cdot \sqrt{1 - \zeta_x^2}
\end{align}
\noindent
and for the combination of parameters relevant for Eq.~\ref{eq.final}:

\begin{align}
  (1 + \cos^2\vartheta)/(\sin\vartheta \cdot L_t) &=  \frac{2 + A^2 \cdot (1 + \zeta_x^2) - 4A \zeta_x}{L \cdot \sqrt{1 - \zeta_x^2} \cdot (1 + A^2 - 2A\zeta_x)}  \quad,\nonumber\\
  &=  \frac{2\cdot (1 + A^2 - 2A \zeta_x) - A^2\cdot (1-\zeta_x^2)}{L \cdot \sqrt{1 - \zeta_x^2} \cdot (1 + A^2 - 2A\zeta_x)}  \quad,\nonumber\\
  &=  \frac{2}{L \cdot \sqrt{1 - \zeta_x^2}} - \frac{A^2 \cdot \sqrt{1-\zeta_x^2}}{L \cdot (1 + A^2 - 2A\zeta_x)}  \label{eq.fscat} \quad.
\end{align}

\section{Derivation of the critical altitude}
\label{app:Hmax}

We assume molecular scattering only, and a critical photo-electron rate $\rcrit$, above which observations are deteriorated.
The condition $\rpix < \rcrit$ yields then a condition for $\Hm$, if the laser points to the direction $(\thetal, \phil)$.

Using Eqs.~\ref{eq.final} and~\ref{eq.fscat}, we obtain:  

\begin{align}
  \rcrit & > (\re{1.4} \cdot 10^{13} ~\mathrm{m}^{-1} \,\mathrm{s}^{-1}) \cdot \nlas \cdot  \pde  \cdot \fovp\cdot \atel \cdot \nonumber\\
  & \quad e^{-\Hm/\Hmol} \cdot \Big(  \frac{2}{L \cdot \sqrt{1 - \zeta_x^2}} - \nonumber\\
  & \qquad \frac{L\cos^2\thetal \cdot \sqrt{1-\zeta_x^2}}{\Hm^2 + L^2\cos^2\thetal - 2L\Hm\cos\thetal\zeta_x)} \Big)  \quad. \label{eq.rcrit}
\end{align}

\begin{figure}
\centering
\includegraphics[width=0.45\linewidth]{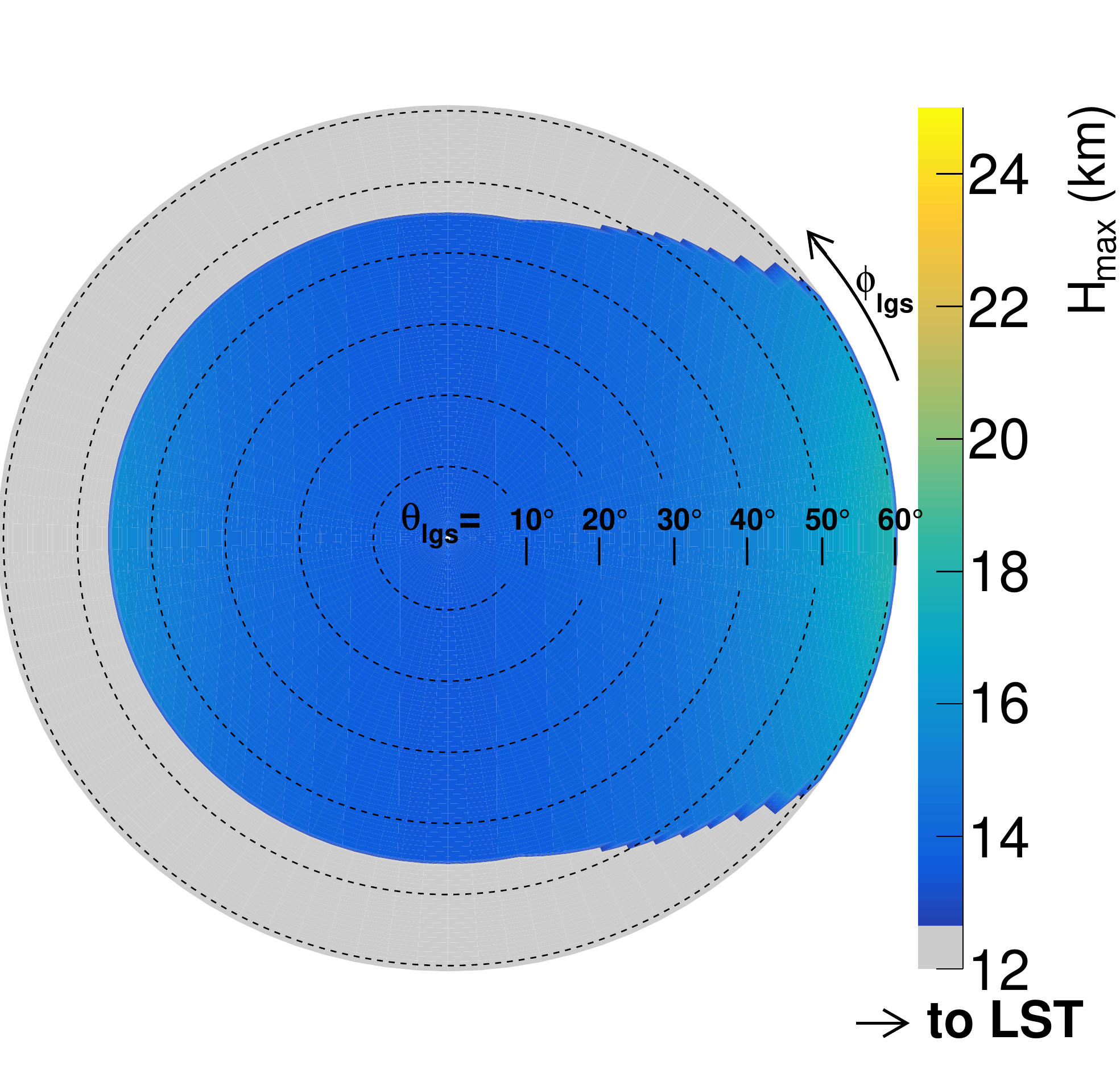}
\includegraphics[width=0.45\linewidth]{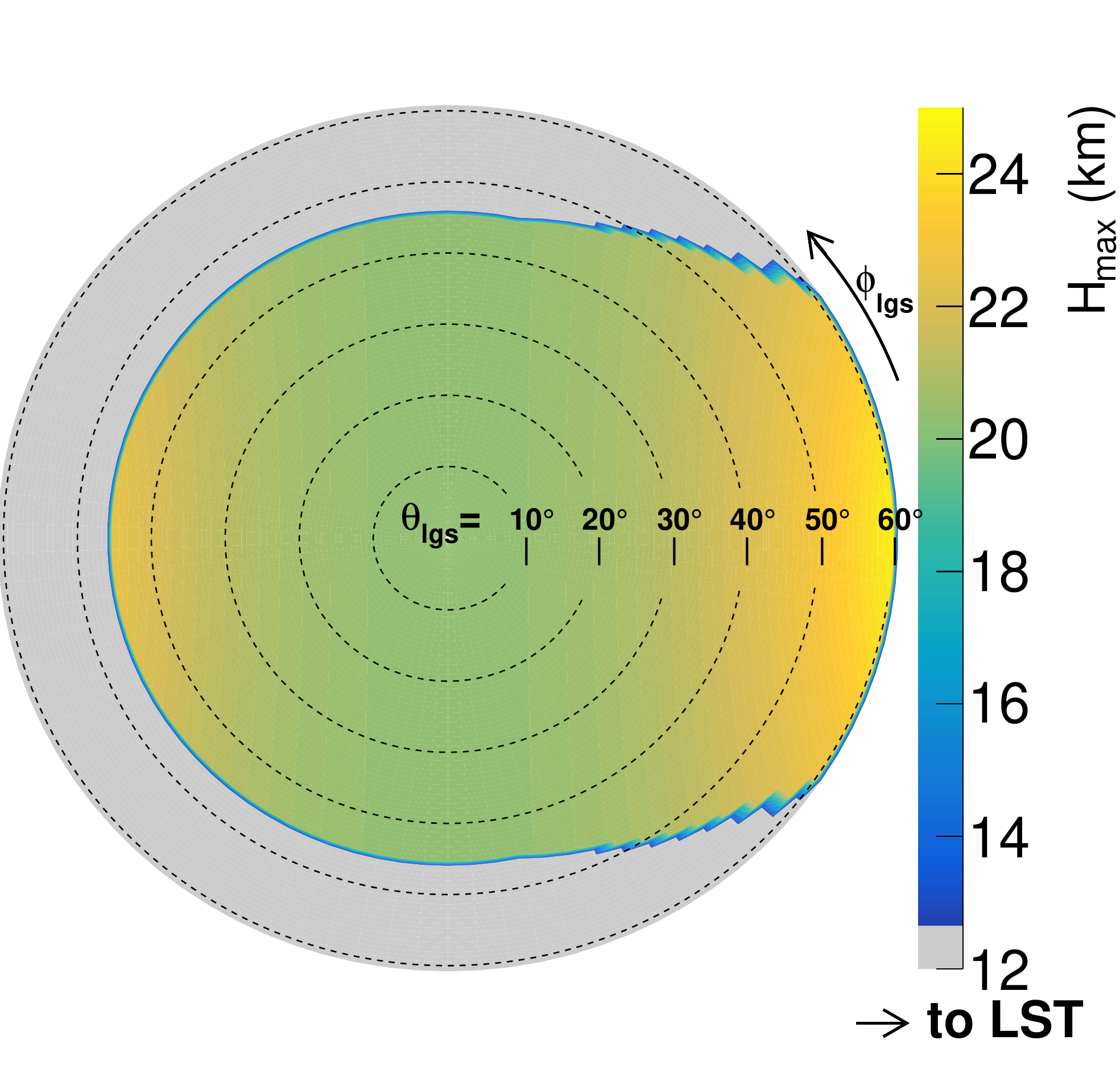}
\caption{Polar plot of the critical altitude for the case of the GTC \re{(left) and the TMT (right)} laser shining into the observability cone of an LST. 
The laser is located in the center, the direction to the LST points towards its right side.
 \label{fig.Hmax}} 
\end{figure}

\begin{figure}
\centering
\includegraphics[width=0.45\linewidth]{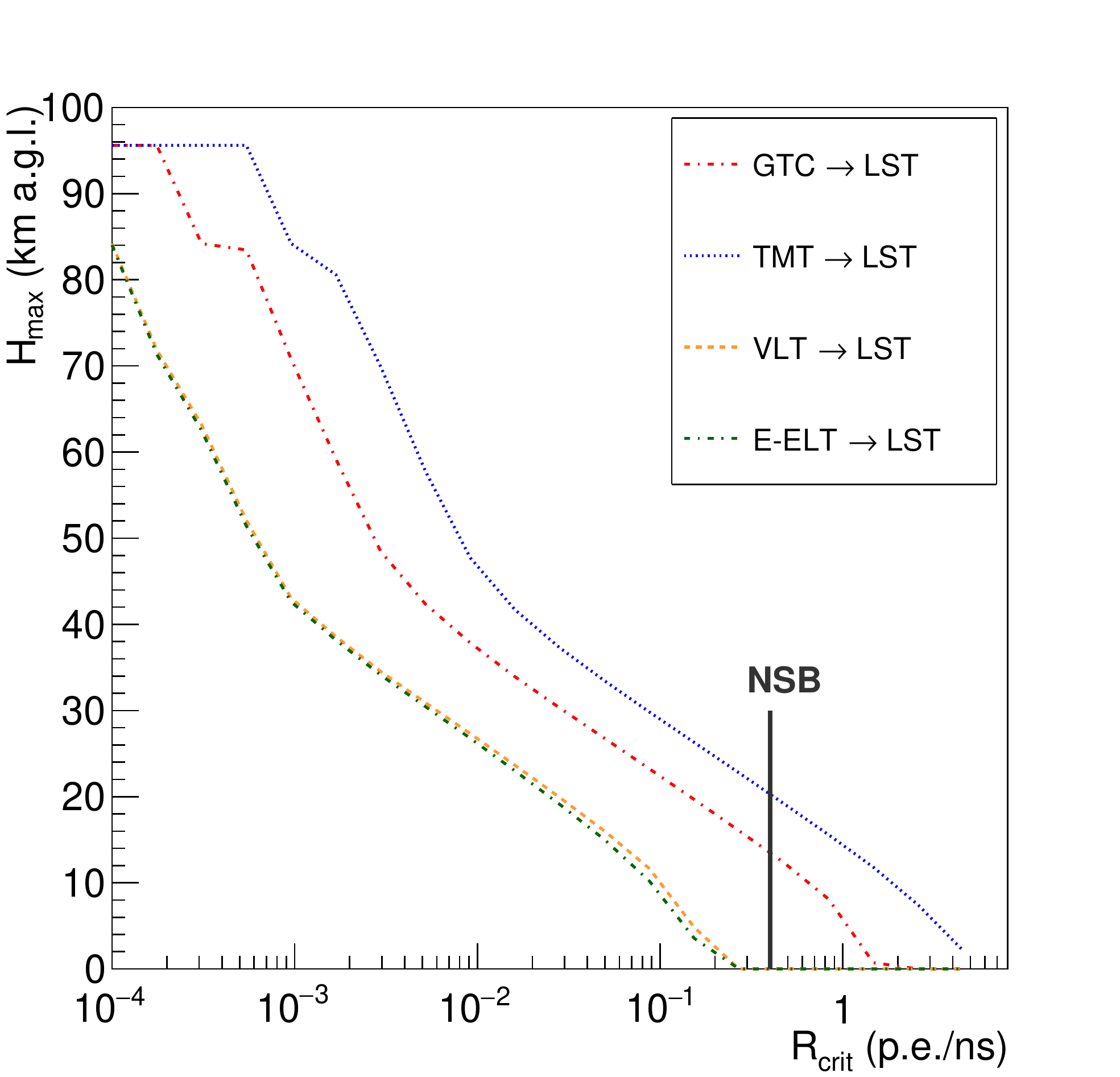}
\caption{Critical altitude $\Hm$ as a function of the critical pixel rate $\rcrit$ for the four investigated LGS systems pointing vertically upwards.
 \label{fig.Hmax_vs_rcrit}} 
\end{figure}

Setting both sides equal, Eq.~\ref{eq.rcrit} can be solved numerically for $\Hm$ using a given combination of $(\thetal, \phil)$.
Figure~\ref{fig.Hmax} shows an example of the critical altitude as a function of the pointing coordinates $(\thetal, \phil)$ of the GTC \re{and the TMT} laser
\re{and Figure~\ref{fig.Hmax_vs_rcrit} as a function of $\rcrit$ for the case of a vertically upward pointing LGS, observed by an LST.}

\section{Derivation of the observability condition and height of intersection point}
\label{sec.derivobs}

We apply Pythagoras' theorem on the triangle contained between the CTA telescope, the projected beam intersection point on ground, and the line connecting the distance between CTA telescope and AO-laser with the intersection point:

\begin{align}
  h^2 \tan^2\theta_\mathrm{CTA} &= y^2 + (L-x)^2  \quad, \nonumber\\
  &= h^2 \cdot \psi_y^2 + ( L - h \psi_x)^2    \quad, \nonumber\\
  &= h^2 \cdot \tan^2\thetal - 2 L\cdot h \psi_x + L^2 \quad. \label{eq.Hm2}       
\end{align}
Requiring that $\theta_\mathrm{CTA} < \thetam$ cannot be larger than a certain maximum zenith angle, and $h<\Hm$, we obtain:
\begin{equation}
\tan^2\thetal - 2 \frac{L}{\Hm} \psi_x + (\frac{L}{\Hm})^2  <  \tan^2\thetam  \quad. \label{eq.theta2thetal}
\end{equation}
\noindent
Note that \autoref{eq.theta2thetal} is also valid for the cases $\psi_x < 0$ and $\Hm\psi_y > L$.
\noindent
Now, we solve \autoref{eq.Hm2} for $h$:
\begin{align}
  \tan^2\thetam  &= \tan^2\thetal - 2 \frac{L}{h} \psi_x + \left(\frac{L}{h}\right)^2  \quad, \nonumber\\
  &= \left( \frac{L}{h} - \psi_x \right)^2 + \psi^2_y \quad, \nonumber\\
 h &= \frac{L}{\psi_x + \sqrt{\tan^2\thetam - \psi_y^2}} \quad. \label{eq.h}       
\end{align}

\vspace{0.5cm}
\section{Derivation of the angular length $\Psi$}
\label{app.psi}

We define, as previously:
\begin{align}
  A  &= L \cdot \cos\thetal / \Hm \quad,  \\
  B  &= \psi_x + \sqrt{\tan^2\thetam - \psi_y^2} \quad.
  \end{align}
Applying the cosine rule for $\Psi$, we obtain:
\begin{equation}
        \cos(\Psi) = \frac{\Dmax^2 + \Dmin^2 - D^2}{2 \Dmax \Dmin} \quad.
\end{equation}
Following the relations:
\begin{align}
       \Dmax  &= \frac{\Hm}{\cos\theta_1} \quad, \nonumber\\
       \Dmin  &= \frac{h}{\cos\thetam}  = \frac{L}{B \cos\thetam} \quad, \nonumber\\
       \Delta &= \frac{\Hm - h}{\cos\thetal} = \frac{1}{\cos\thetal} \cdot (\Hm - \frac{L}{B})  \quad,
\end{align}
\noindent
we obtain:
\begin{align}
  \cos(\Psi) &= \frac{1}{2} \cdot \Bigg\{\frac{\Hm}{h} \cdot  \frac{\cos\thetam}{\cos\theta_1} + \frac{h}{\Hm} \cdot  \frac{\cos\theta_1}{\cos\thetam} 
                     - \frac{\cos\theta_1\cos\thetam}{\cos^2\thetal} \cdot \left(\frac{h}{\Hm} +  \frac{\Hm}{h} - 2 \right) \Bigg\} \\[0.3cm]
                     & \mathrm{with:}  \quad   \frac{1}{\cos^2_1} =   \frac{1}{\cos^2\thetal} - 2\frac{L}{\Hm}\psi_x + \left(\frac{L}{\Hm}\right)^2 \qquad, \nonumber\\[0.3cm]                    
                     &  \qquad \qquad  ~\frac{1}{\cos_1}          =   \frac{1}{\cos\thetal} \cdot \sqrt{1 + A^2 - 2A\zeta_x} \quad,\\[0.3cm]
                     &  \qquad \qquad \frac{\Hm}{h}               =  \frac{\Hm}{L} \cdot  B  \quad,\quad \mathrm{with~\Hm~obtained~from~Equation~\ref{eq.rcrit},}
\end{align}
\noindent
where the latter can easily derived from \autoref{eq.Hm2}.
\begin{align}
\end{align}

\begin{align}
  \cos(\Psi) &= \frac{1}{2} \cdot  \frac{B\cdot\cos\thetam\cdot (1 + A^2 - 2A\zeta_x) + A^2/(B\cdot \cos\thetam) - \cos\thetam/B \cdot (B-A/\cos\thetal)^2}{A\cdot \sqrt{1+A^2-2A\zeta_x}} \quad, \nonumber\\[0.3cm]
  &= \frac{\cos\thetam}{2} \cdot  \frac{B\cdot (A - 2\zeta_x) + \frac{A}{B} \cdot  (\frac{1}{\cos^2\thetam} - \frac{1}{\cos^2\thetal}) + \frac{2}{\cos\thetal} }{\sqrt{1+A^2-2A\zeta_x}} \quad.  \nonumber\\[0.3cm]
\end{align}

\end{appendix}

\end{document}